\documentclass[oldversion]{aa}

\usepackage{graphicx}
\usepackage{txfonts}
\usepackage{lscape}
\usepackage{ulem}

\begin{document}
   \title{MAXI\,J1659$-$152: The shortest orbital period black-hole transient in outburst}

   \author{E. Kuulkers\inst{1}
	  \and
	  C.\ Kouveliotou\inst{2}
	  \and
	  T.\ Belloni\inst{3}
	  \and
	  M.\ Cadolle Bel\inst{1}
	  \and
	  J. Chenevez\inst{4}
	  \and
	  M.\ D\'\i az Trigo\inst{5}
	  \and
	  J.\ Homan\inst{6}
	  \and
	  A.\ Ibarra\inst{1}
	  \and
	  J.A.\ Kennea\inst{7}
	  \and
	  T.\ Mu\~noz-Darias\inst{3,8} 
	  \and
	  J.-U.\ Ness\inst{1}
	  \and
	  A.N.\ Parmar\inst{1}
	  \and
	  A.M.T.\ Pollock\inst{1}
	  \and
	  E.P.J.\ van den Heuvel\inst{9}
	  \and
	  A.J.\ van der Horst\inst{9,10}
	  }

   \authorrunning{E.~Kuulkers et al.}
   \titlerunning{MAXI\,J1659$-$152}

   \offprints{E.~Kuulkers}

   \institute{European Space Astronomy Centre (ESA/ESAC), Science Operations Department, 28691 Villanueva de la Ca\~nada (Madrid), Spain \\
              \email{Erik.Kuulkers@esa.int}
          \and
	      Astrophysics Office, ZP12, NASA/Marshall Space Flight Center, Huntsville, AL 35812, USA
          \and
	      INAF-Osservatorio Astronomico di Brera, Via E. Bianchi 46, I-23807 Merate (LC), Italy
	  \and 
	      National Space Institute, Technical University of Denmark, Juliane Maries Vej 30, 2100 Copenhagen, Denmark
          \and
	      ESO, Karl-Schwarzschild-Strasse 2, 85748 Garching bei M\"unchen, Germany
          \and
	      MIT Kavli Institute for Astrophysics and Space Research, 70 Vassar Street, Cambridge, MA 02139, USA
          \and
              Department of Astronomy \&\ Astrophysics, The Pennsylvania State University, 525 Davey Lab, University Park, PA 16802, USA
          \and
	      School of Physics and Astronomy, University of Southampton, SO17 1BJ, United Kingdom
          \and
	      Astronomical Institute `Anton Pannekoek', University of Amsterdam, Science Park 904, 1098 XH Amsterdam, The Netherlands
          \and
	      Universities Space Research Association, NSSTC, 320 Sparkman Drive, Huntsville, AL 35805, USA
             }

   \date{Received; accepted}

\abstract{
MAXI\,J1659$-$152 is a bright X-ray transient black-hole candidate binary system discovered in September 2010.
We report here on MAXI, RXTE, {\it Swift}, and {\it XMM-Newton} observations during its 2010/2011 outburst.
We find that during the first one and a half week of the outburst the X-ray light curves display drops in intensity at 
regular intervals, which we interpret as absorption dips. About three weeks into the outbursts, again drops in intensity
are seen. These dips have, however, a spectral behaviour opposite to that of the absorption dips, and are related to fast 
spectral state changes (hence referred to as transition dips).
The absorption dips recur with a period of 2.414$\pm$0.005\,hrs, which we interpret as the orbital
period of the system. This implies that MAXI\,J1659$-$152 is the shortest period black-hole candidate binary known to date.
The inclination of the accretion disk with respect to the
line of sight is estimated to be 65--80$^{\circ}$. We propose the companion to the black-hole candidate to be close to an 
M5 dwarf star, with a mass and radius of about 0.15--0.25\,M$_{\odot}$ and 0.2--0.25\,R$_{\odot}$, respectively. 
We derive that the companion had an initial mass of about 1.5\,M$_{\odot}$, which evolved
to its current mass in about 5--6 billion years.
The system is rather compact (orbital separation of $\gtrsim$1.33\,R$_{\odot}$), and is located at a distance of 8.6$\pm$3.7\,kpc,
with a height above the Galactic plane of 2.4$\pm$1.0\,kpc.
The characteristics of short orbital period and high Galactic scale height are shared with two other transient 
black-hole candidate X-ray binaries, i.e., XTE\,J1118+480 and Swift\,J1735.5$-$0127. 
We suggest that all three are kicked out of the Galactic plane into the halo, rather than being formed in a globular cluster.
}


   \maketitle
%

\section{Introduction}
\label{intro}

Transient X-ray sources have been extensively studied since the advent of X-ray astronomy. Most are in binary systems comprising at least one compact object, 
which is either a neutron star or a black hole. Determining the nature of the compact object is not trivial, as a number of their X-ray characteristics are 
common to both types of objects. Black hole X-ray binaries (BHXBs) are uniquely identified by a mass determination of the compact object in excess of 
3\,M$_{\odot}$. The first two confirmed BHXBs were Cyg\,X-1 (Webster \&\ Murdin 1972, Bolton 1972) 
and LMC\,X-3 (Cowley et al.\ 1983); these systems were, however, persistent X-ray sources. The very first transient 
source with a confirmed black hole, A0620$-$00 (McClintock \&\ Remillard 1986), was discovered in 1975, when it reached intensities of $\sim$50\,Crab (Elvis et al.\ 1975). To date there are 
20 known BHXBs, of which 17 are known to be transient (e.g., Remillard \&\ McClintock 2006).
The BHXB transients belong to the class of low-mass X-ray binaries (LMXBs), i.e., binaries containing a low-mass (typically $\lesssim$1\,M$_{\odot}$) 
companion orbiting a neutron star or a black hole.

BHXBs are primarily discovered when they enter outbursts characterised by increased X-ray luminosities, by a factor of at least 100 in a few days, and a variety of spectral 
and temporal variability states. The most contrasted states are the so-called `hard' and the `soft' states. The former, originally was also defined as a `low' state, 
owing to the representation of the spectrum by a power law with spectral index 1.4--2.0 up to several hundreds of keV. Conversely, the latter `soft', or `high', state 
is characterised by a tenfold increase of the $\sim$2--10\,keV flux. Several other states have been defined in the literature, depending also on the evolution of the 
combined spectral and temporal variability properties (for extensive reviews, see Homan \&\ Belloni 2005, Belloni 2010, and 
McClintock \&\ Remillard 2006, Remillard \&\ McClintock 2006).

On 2010 September 25 08:05 UTC, the {\it Swift}/Burst Alert Telescope (BAT) triggered on a source located roughly 17$^{\circ}$ above the Galactic plane. 
The source was initially designated as Gamma-Ray Burst (GRB) 100925A (Mangano et al.\ 2010), and was monitored with the {\it Swift}/X-Ray Telescope (XRT). Interestingly, 
the source flux did not decline during the next several hours, as is usually the trend with GRBs. This unusual behaviour and the source location near the 
Galactic bulge indicated that it might not be a GRB but a new Galactic source (Kahn 2010). Later that day, the {\it Monitor of All-sky X-ray Image} (MAXI) team 
reported the detection of a new hard X-ray transient, MAXI\,J1659$-$152, whose position was consistent with GRB\,100925A and which had brightened since 
2010 September 25 02:30 UTC (Negoro et al.\ 2010). The discovery initiated several multi-wavelength observations 
(ground and space-based; e.g., van der Horst et al.\ 2010, Kuulkers et al.\ 2010b, 2010c; see also Kuulkers et al.\ 2012a).

The Galactic origin of MAXI\,J1659$-$152 was also suggested the next day from a combined UV-X-ray spectral-energy distribution analysis 
using data from the {\it Swift}/XRT 
and {\it Swift}/UV Optical Telescope (UVOT) (Xu 2010). Finally, optical spectroscopy data taken with the ESO/Very Large Telescope (VLT) X-shooter instrument showed 
various broad emission lines from the Balmer and Paschen series of H and of He\,II, as well as Ca\,II and Na\,I absorption lines from the interstellar medium, all at redshift zero, 
clinching the Galactic nature of the source. Moreover, the double-peaked profiles of the emission lines indicated that the source was an X-ray binary 
(de Ugarte Postigo et al.\ 2010, Kaur et al.\ 2012).
The binary nature of the source has by now been established in several studies of its spectral and temporal properties. 
Kennea et al.\ (2010) reported frequent intensity drops in the X-ray light curve, attributed possibly to eclipses by a companion star. 
Kuulkers et al.\ (2010d, 2012a) established a period of $\sim$2.42\,hr using the {\it X-ray Multi Mirror mission (XMM-Newton)} data, which was later confirmed by Belloni et al.\ (2010, {\it Rossi} X-ray Timing Explorer (RXTE); 
also Kuulkers et al.\ 2012a) and Kennea et al.\ (2011; {\it Swift}). 
Kuroda et al.\ (2010) have reported optical variations up to 0.1\,mag, consistent with a double-peaked modulation at a period of 2.4158$\pm$0.0003\,hrs.
We concluded that this is the shortest BHXB orbital period measured as yet (Kuulkers et al.\ 2010d, 2012a), and suggested
that the source was most likely viewed at a high inclination (Kuulkers et al.\ 2012a).

The source was identified as a BHXB through the fast timing behaviour observed with the 
RXTE/Proportional Counter Array (PCA), which was similar to that seen in stellar-mass black-hole transients 
(Kalamkar et al.\ 2011, Kennea et al.\ 2011, Mu\~noz-Darias et al.\ 2011b, Shaposhnikov et al.\ 2012, Yamaoka et al.\ 2012). 
Estimates of the mass of the compact object range from 2.2--3.1\,M$_{\odot}$ (Kennea et al.\ 2011), to 3.6--8.0\,M$_{\odot}$ (Yamaoka et al.\ 2012), to
20$\pm$3\,M$_{\odot}$ (Shaposhnikov et al.\ 2012). Part of the discrepancy can be resolved by taking into account the spin of the black hole 
(Kennea et al.\ 2011, Yamaoka et al.\ 2012).
The source distance estimates range between 1.6$-$4.2\,kpc (Miller-Jones et al.\ 2011) and 8.6\,kpc (Yamaoka et al.\ 2011). 

After the main outburst, MAXI\,J1659$-$152 continued to show low-level activity (e.g., Kennea et al.\ 2011).
The flux of MAXI\,J1659$-$152 was observed to exhibit sudden intensity changes, e.g., during 2011 May, 
when {\it Swift} 
and {\it Chandra} detected a reflare. The source then decayed again over a period of $\sim$2 months
(Yang \&\ Wijnands 2011a, 2011b, 2011c, Jonker et al.\ 2012). It seemed to enter its quiescent state during the second half of
2011 (Yang \&\ Wijnands 2011d, Russell et al.\ 2011, Kong et al.\ 2011).
However, for a BHXB it was still too bright to be in true quiescence (Jonker et al.\ 2012).
A possible quiescent optical counterpart has been reported based on observations done on 2010 June 19 (about 3 months before 
the start of the outburst) and 2012 March 23 (Kong et al.\ 2010, Kong 2012).

Our team was involved in various Target of Opportunity (ToO) observations taken with 
RXTE, {\it Swift} and {\it XMM-Newton}.
Here we present a detailed account of the X-ray and UV light curves of MAXI\,J1659$-$152, taken over the course of the 2010 outburst with 
these satellites and with MAXI.
A preliminary account of the results can be found in Kuulkers et al.\ (2010b, 2010d, 2012a) and Belloni et al.\ (2010). 
The MAXI, RXTE and {\it Swift} data have already been (partly) described elsewhere, however, the focus was more on the combined spectral and 
timing behaviour of MAXI\,J1659$-$152 (Kalamkar et al.\ 2011, Kennea et al.\ 2011, Mu\~noz-Darias et al.\ 2011b, Shaposhnikov et al.\ 2012, Yamaoka et al.\ 2012).
We find that MAXI\,J1659$-$152 is indeed viewed at a high inclination. It still has the shortest 
orbital period known to date, and is, therefore, a rather compact BHXB transient. 
We present in Sect.~2 a detailed description of the data sets used in our analysis, and in Sect.~3 the results of our timing analysis 
of the variations of the source light curve. We discuss the observed light curve dips in Sect.~4 and provide an interpretation on the characteristics 
of the binary system and its distance. 

\section{Observations}
\label{data_analysis}

\begin{table*}
\caption{Log of X-ray observations with RXTE, {\it Swift} and {\it XMM-Newton} during the main outburst of MAXI\,J1659$-$152 presented in this paper, ordered along the start time of the observation.}
\begin{tabular}{rcrcc|rcrcc}
\hline
\multicolumn{1}{c}{Day$^1$} &
\multicolumn{1}{c}{Start time (UT)} &
\multicolumn{1}{c}{Exp.$^2$} &
\multicolumn{1}{c}{ObsID} &
\multicolumn{1}{c|}{Satellite} &
\multicolumn{1}{c}{Day$^1$} &
\multicolumn{1}{c}{Start time (UT)} &
\multicolumn{1}{c}{Exp.$^2$} &
\multicolumn{1}{c}{ObsID} &
\multicolumn{1}{c}{Satellite} \\
\hline
 0.3 & 2010-09-25 07:49 & 19667 & 00434928000    & Swift      & 16.6 & 2010-10-11 15:30 &  2028 & 95108-01-25-00 & RXTE       \\
 1.0 & 2010-09-26 00:07 & 16214 & 00434928001    & Swift      & 17.0 & 2010-10-12 00:36 &  3348 & 95108-01-26-00 & RXTE       \\
 1.6 & 2010-09-26 13:31 &  9909 & 00434928002    & Swift      & 17.7 & 2010-10-12 16:27 &  2453 & 95108-01-27-00 & RXTE       \\
 2.0 & 2010-09-27 00:13 & 19108 & 00434928003    & Swift      & 18.4 & 2010-10-13 09:43 &  2697 & 95108-01-28-00 & RXTE       \\
 2.7 & 2010-09-27 16:15 & 51916 & 0656780601     & XMM-Newton & 18.8 & 2010-10-13 19:24 &  1284 & 00434928023    & Swift      \\
 3.0 & 2010-09-28 00:53 & 16910 & 95358-01-02-00 & RXTE       & 19.5 & 2010-10-14 11:37 &  1705 & 00434928025    & Swift      \\
 3.3 & 2010-09-28 07:06 & 10009 & 00434928005    & Swift      & 19.9 & 2010-10-14 21:32 &  3516 & 95108-01-30-00 & RXTE       \\
 4.1 & 2010-09-29 01:58 &  1706 & 95358-01-02-01 & RXTE       & 20.0 & 2010-10-15 00:28 &  1240 & 00434928026    & Swift      \\
 4.2 & 2010-09-29 05:14 &  2318 & 00434928007    & Swift      & 20.2 & 2010-10-15 05:27 &  3444 & 95118-01-01-00 & RXTE       \\
 5.1 & 2010-09-30 02:00 &  1410 & 95358-01-02-02 & RXTE       & 20.6 & 2010-10-15 13:27 &  1390 & 00434928027    & Swift      \\
 5.2 & 2010-09-30 05:37 &  2719 & 00434928008    & Swift      & 20.7 & 2010-10-15 16:24 &  3554 & 95118-01-01-01 & RXTE       \\
 6.2 & 2010-10-01 05:45 &  3369 & 95358-01-03-00 & RXTE       & 21.0 & 2010-10-16 00:34 &  1305 & 00434928028    & Swift      \\
 6.2 & 2010-10-01 05:45 &  2594 & 00434928009    & Swift      & 21.1 & 2010-10-16 03:25 &  3374 & 95118-01-02-00 & RXTE       \\
 6.4 & 2010-10-01 10:44 &  2130 & 95108-01-02-00 & RXTE       & 21.5 & 2010-10-16 11:45 &  1325 & 00434928029    & Swift      \\
 7.1 & 2010-10-02 02:31 &  1763 & 95358-01-03-01 & RXTE       & 22.0 & 2010-10-17 00:38 &  1390 & 00434928030    & Swift      \\
 7.1 & 2010-10-02 02:37 &  2369 & 00434928010    & Swift      & 22.2 & 2010-10-17 04:25 &   829 & 95118-01-03-01 & RXTE       \\
 7.5 & 2010-10-02 11:50 &  1905 & 95108-01-03-00 & RXTE       & 22.8 & 2010-10-17 18:55 &  2464 & 95118-01-03-00 & RXTE       \\
 7.8 & 2010-10-02 18:22 &  1175 & 95108-01-04-00 & RXTE       & 23.0 & 2010-10-18 00:46 &  1289 & 00434928032    & Swift      \\
 8.1 & 2010-10-03 01:41 &   994 & 95108-01-05-00 & RXTE       & 23.7 & 2010-10-18 16:36 &  3556 & 95118-01-04-00 & RXTE       \\
 8.1 & 2010-10-03 02:36 &  1720 & 00434928011    & Swift      & 24.0 & 2010-10-19 00:24 &  3027 & 95118-01-05-00 & RXTE       \\
 8.2 & 2010-10-03 04:50 &  3352 & 95358-01-03-02 & RXTE       & 24.0 & 2010-10-19 00:50 &  1290 & 00031843001    & Swift      \\
 8.5 & 2010-10-03 11:20 &  1737 & 95108-01-06-00 & RXTE       & 24.6 & 2010-10-19 13:31 &  1245 & 00031843002    & Swift      \\
 8.9 & 2010-10-03 21:03 &  1378 & 95108-01-07-00 & RXTE       & 24.9 & 2010-10-19 20:51 &  2358 & 95118-01-05-01 & RXTE       \\
 9.1 & 2010-10-04 02:40 &  3314 & 00434928012    & Swift      & 25.0 & 2010-10-20 00:56 &  1290 & 00031843003    & Swift      \\
 9.1 & 2010-10-04 02:43 &  3352 & 95108-01-08-00 & RXTE       & 25.3 & 2010-10-20 06:17 &  3495 & 95118-01-06-00 & RXTE       \\
 9.5 & 2010-10-04 10:52 &  2216 & 95108-01-09-00 & RXTE       & 25.5 & 2010-10-20 12:21 &  1089 & 00031843004    & Swift      \\
 9.7 & 2010-10-04 17:19 &  1309 & 95108-01-10-00 & RXTE       & 25.7 & 2010-10-20 17:16 &  3468 & 95118-01-06-01 & RXTE       \\
10.1 & 2010-10-05 02:47 &  3414 & 00434928013    & Swift      & 26.0 & 2010-10-21 00:53 &   985 & 00031843005    & Swift      \\
10.6 & 2010-10-05 13:43 &  1847 & 95108-01-11-00 & RXTE       & 26.1 & 2010-10-21 02:40 &  2883 & 95118-01-07-01 & RXTE       \\
10.8 & 2010-10-05 20:02 &  1676 & 95108-01-12-00 & RXTE       & 26.6 & 2010-10-21 13:46 &   985 & 00031843006    & Swift      \\
11.1 & 2010-10-06 02:53 &  3284 & 00434928014    & Swift      & 26.7 & 2010-10-21 16:48 &  3214 & 95118-01-07-00 & RXTE       \\
11.4 & 2010-10-06 09:48 &  2397 & 95108-01-13-00 & RXTE       & 27.0 & 2010-10-22 00:37 &   893 & 95118-01-08-00 & RXTE       \\
11.7 & 2010-10-06 17:55 &  1443 & 95108-01-14-00 & RXTE       & 27.0 & 2010-10-22 00:57 &  1264 & 00031843007    & Swift      \\
12.1 & 2010-10-07 01:17 &  3409 & 95108-01-15-00 & RXTE       & 27.2 & 2010-10-22 05:51 &  1040 & 00031843008    & Swift      \\
12.1 & 2010-10-07 02:47 &  1575 & 00434928015    & Swift      & 27.8 & 2010-10-22 19:22 &  1940 & 95118-01-09-00 & RXTE       \\
12.4 & 2010-10-07 09:28 &  2337 & 95108-01-16-00 & RXTE       & 29.2 & 2010-10-24 05:54 &  1162 & 95118-01-10-00 & RXTE       \\
12.7 & 2010-10-07 15:38 &  1595 & 00434928016    & Swift      & 30.2 & 2010-10-25 05:24 &   888 & 95118-01-11-00 & RXTE       \\
12.7 & 2010-10-07 15:47 &  1602 & 95108-01-17-00 & RXTE       & 31.0 & 2010-10-26 00:39 &   868 & 95118-01-12-00 & RXTE       \\
13.0 & 2010-10-07 23:37 &   858 & 95108-01-18-00 & RXTE       & 32.5 & 2010-10-27 12:25 &  2197 & 95118-01-13-00 & RXTE       \\
13.0 & 2010-10-08 00:00 &  1015 & 95108-01-18-01 & RXTE       & 33.5 & 2010-10-28 11:54 &   774 & 95118-01-14-00 & RXTE       \\
13.1 & 2010-10-08 02:52 &  1630 & 00434928017    & Swift      & 34.5 & 2010-10-29 11:31 &  1152 & 95118-01-15-00 & RXTE       \\
13.7 & 2010-10-08 16:55 &  1662 & 95108-01-19-00 & RXTE       & 35.3 & 2010-10-30 07:54 &  1412 & 95118-01-15-01 & RXTE       \\
14.0 & 2010-10-09 00:49 &  1900 & 95108-01-20-00 & RXTE       & 36.3 & 2010-10-31 07:24 &  1458 & 95118-01-16-00 & RXTE       \\
14.1 & 2010-10-09 02:58 &  1584 & 00434928019    & Swift      & 37.2 & 2010-11-01 05:20 &  1467 & 95118-01-16-01 & RXTE       \\
14.5 & 2010-10-09 11:39 &  2309 & 95108-01-21-00 & RXTE       & 38.0 & 2010-11-02 00:14 &  1715 & 95118-01-17-00 & RXTE       \\
14.7 & 2010-10-09 15:49 &  1595 & 00434928020    & Swift      & 39.0 & 2010-11-03 01:11 &  1457 & 95118-01-17-01 & RXTE       \\
15.1 & 2010-10-10 01:27 &  1009 & 00434928021    & Swift      & 40.1 & 2010-11-04 01:15 &   786 & 95118-01-18-00 & RXTE       \\
15.1 & 2010-10-10 03:05 &  3507 & 95108-01-22-00 & RXTE       & 41.0 & 2010-11-05 00:16 &  2348 & 95118-01-19-00 & RXTE       \\
15.5 & 2010-10-10 12:42 &  1629 & 00434928022    & Swift      & 42.2 & 2010-11-06 04:32 &  1139 & 95118-01-20-00 & RXTE       \\
15.7 & 2010-10-10 15:54 &  1914 & 95108-01-23-00 & RXTE       & 44.1 & 2010-11-08 01:58 &  1608 & 95118-01-21-00 & RXTE       \\
16.2 & 2010-10-11 04:03 &  3513 & 95108-01-24-00 & RXTE       &  &  &  &  &  \\                                                 
\hline
\end{tabular}
\note{Day = MJD - 55464, where MJD\,55464 corresponds to UT 2010, September 25, 0:00.}
\note{Total on-source exposure time in sec.}
\label{xraylog}
\end{table*}

In this paper we only concentrate on the light curves of the various instruments 
onboard MAXI (Matsuoka et al. 2009), RXTE (Bradt et al.\ 1993), {\it Swift} (Gehrels et al.\ 2004) 
and {\it XMM-Newton}\footnote{We
note that the {\it XMM-Newton} observations were taken simultaneously with one of the INTEGRAL ToO observations of the source, see 
Kuulkers et al.\ (2012a).} (Jansen et al.\ 2001). 
We refer to Table~1 for an observation log of the latter 3 satellites.
The {\it XMM-Newton} spectral analysis is 
deferred to a future paper. For a spectral analysis of the RXTE and {\it Swift} data we refer to Mu\~noz-Darias et al.\ (2011b) 
and Kennea et al.\ (2011), respectively, as well as Yamaoka et al.\ (2012).
All our light curves have been subjected to a barycentric correction using the standard tools available.

\subsection{XMM-Newton}
\label{XMM}

We used Science Analysis System (SAS) version 11.0.0 together with the latest calibration
files to analyse the {\it XMM-Newton} data. The European Photon Imaging Camera (EPIC)-MOS (Turner et al.\ 2001)
cameras were not used during the observation in order to allocate
their telemetry to the EPIC-pn camera (Str\"uder et al.\ 2001) and to avoid full scientific
buffer in the latter. The EPIC-pn was used in timing mode,
whilst the Reflection Grating Spectrometer instruments (RGS1 and RGS2; den Herder et al.\ 2001) were operating in the
standard spectroscopy mode. Due to the brightness of the source
the EPIC-pn data were affected by pile-up (see below). Similarly, 
the count rate was also above the limits of pile-up for the RGS2,
for which the read-out is slower by a factor of two compared to that of RGS1 since August 2007
(see section 3.4.4.8 of the {\it XMM-Newton} Users Handbook). 

Standard data reduction procedures (SAS tasks {\tt epproc} and
{\tt rgsproc}) were used to obtain EPIC-pn and RGS calibrated event
files. We used the task {\tt epfast} on the event files to correct for charge transfer
inefficiency (CTI) effects seen in the EPIC-pn timing mode
when high count rates are present. 

The count rate in the EPIC-pn was close to, or above, the
800\,cts\,s$^{-1}$ level, at which X-ray loading and pile-up effects
become significant. 
Pile-up occurs when more than one photon is read in a pixel during a
read-out cycle. This causes photon loss, pattern
migration from lower to higher pattern types and hardening of the spectrum,
because the charges deposited by more than one photon are added up before being read 
out (see the {\it XMM-Newton} Users Handbook for more information on pile-up). 

Since pile-up causes significant spectral distortion and a decline in the count rate measured by {\it XMM-Newton},
we investigated in detail its presence before extracting the time series. 
We used the SAS task {\tt epatplot}, which utilises the relative ratios of single- and
double-pixel events which deviate from standard values in case of
significant pile-up, as a diagnostic tool in the pn camera timing mode
data and found that the spectrum was affected by pile-up. Next, we
extracted several spectra selecting single and double timing mode
events (patterns 0 to 4) but different spatial regions for the
source. Source events were first extracted from a 62\arcsec\ (15
columns) wide box centred on the source position (Region~1). Next we
excluded 2,4,6,8 and 10 columns from the centre of Region~1
(Regions~2--6) and extracted one spectrum for each of the defined
regions. We found that the spectrum was free of pile-up after removing 
the central 10 columns.
Then, we used this free of pile-up event list to extract the time series 
shown in this paper.

In the case of the RGS, we used table 11 in the {\it XMM-Newton} Users Handbook 
to determine which CCDs were affected by pile-up. We found that CCDs 6,7 and
8 from RGS2 were above the pile-up limits and, therefore, we did not use them for analysis. 

The EPIC-pn and RGS time series at 1\,s and 100\,s time resolution were produced using the 
{\tt epiclccorr} and {\tt rgslccorr} tasks, respectively. 

In the EPIC-pn timing mode, there are no source-free background
regions, since the point-spread function of the telescope extends further than the
central CCD boundaries. In the case of RGS, since MAXI\,J1659$-$152 was very bright, its
spectrum is not significantly modified by the
`real' background which contributes less than 1\%\ to the total count
rate in most of the bandwidth.  Therefore, we chose not to 
subtract the `background' extracted from 
the outer regions of the central CCD (see also Done \&\ Diaz Trigo 2010, Ng et al.\ 2010). 

The final RGS light curve was calculated by combining
RGS1 and RGS2, adding both orders 1 and 2.

The optical monitor (OM, Talavera 2009, Mason et al.\ 2001) 
was operated in EPIC imaging mode with the two filters UVW1 ($\lambda$$\sim$2500--3500\,\AA) 
and UVM2 ($\lambda$$\sim$2000--2600\,\AA).
Ten consecutive exposures of 4200\,s duration each were taken, first five in the UVW1, and then five more in the UVM2 filter.
We derived the source magnitudes from the output of standard extraction
using the SAS task {\tt omichain}. 
In Table~2, the average fluxes and magnitudes in each filter are listed for each of the exposures.

\begin{table}
\caption{Log of observations with the {\it XMM-Newton}/OM.}
\begin{tabular}{rccc}
\hline
\multicolumn{1}{c}{Time$^3$} &
\multicolumn{1}{c}{Filter} &
\multicolumn{1}{c}{Flux} &
\multicolumn{1}{c}{Magnitude} \\
\multicolumn{1}{c}{(ksec)} &
\multicolumn{1}{c}{} &
\multicolumn{1}{c}{(10$^{-15}$\,erg\,cm$^{-2}$\,s$^{-1}$\,\AA $^{-1}$)} &
\multicolumn{1}{c}{} \\
\hline
0.00 & UVW1 & 1.82 $\pm$ 0.02 & 15.75 $\pm$ 0.02 \\
4.52 & UVW1 & 1.81 $\pm$ 0.02 & 15.76 $\pm$ 0.02 \\
9.05 & UVW1 & 1.79 $\pm$ 0.02 & 15.77 $\pm$ 0.02 \\
13.57 & UVW1 & 1.79 $\pm$ 0.02 & 15.77 $\pm$ 0.02 \\
18.09 & UVW1 & 1.92 $\pm$ 0.03 & 15.69 $\pm$ 0.02 \\
22.61 & UVM2 & 1.18 $\pm$ 0.05 & 16.45 $\pm$ 0.05 \\
27.13 & UVM2 & 1.32 $\pm$ 0.05 & 16.33 $\pm$ 0.04 \\
31.65 & UVM2 & 1.22 $\pm$ 0.05 & 16.41 $\pm$ 0.05 \\
37.97 & UVM2 & 1.35 $\pm$ 0.05 & 16.31 $\pm$ 0.04 \\
42.50 & UVM2 & 0.95 $\pm$ 0.03 & 16.69 $\pm$ 0.03 \\
\hline
\end{tabular}
\note{Start of the exposure in ksec relative to UTC 2010 September 27 16:24:36 (= MJD\,55466.68375).}
\label{omlog}
\end{table}

\begin{figure*}[t!]
\centering
  \includegraphics[height=.79\textheight]{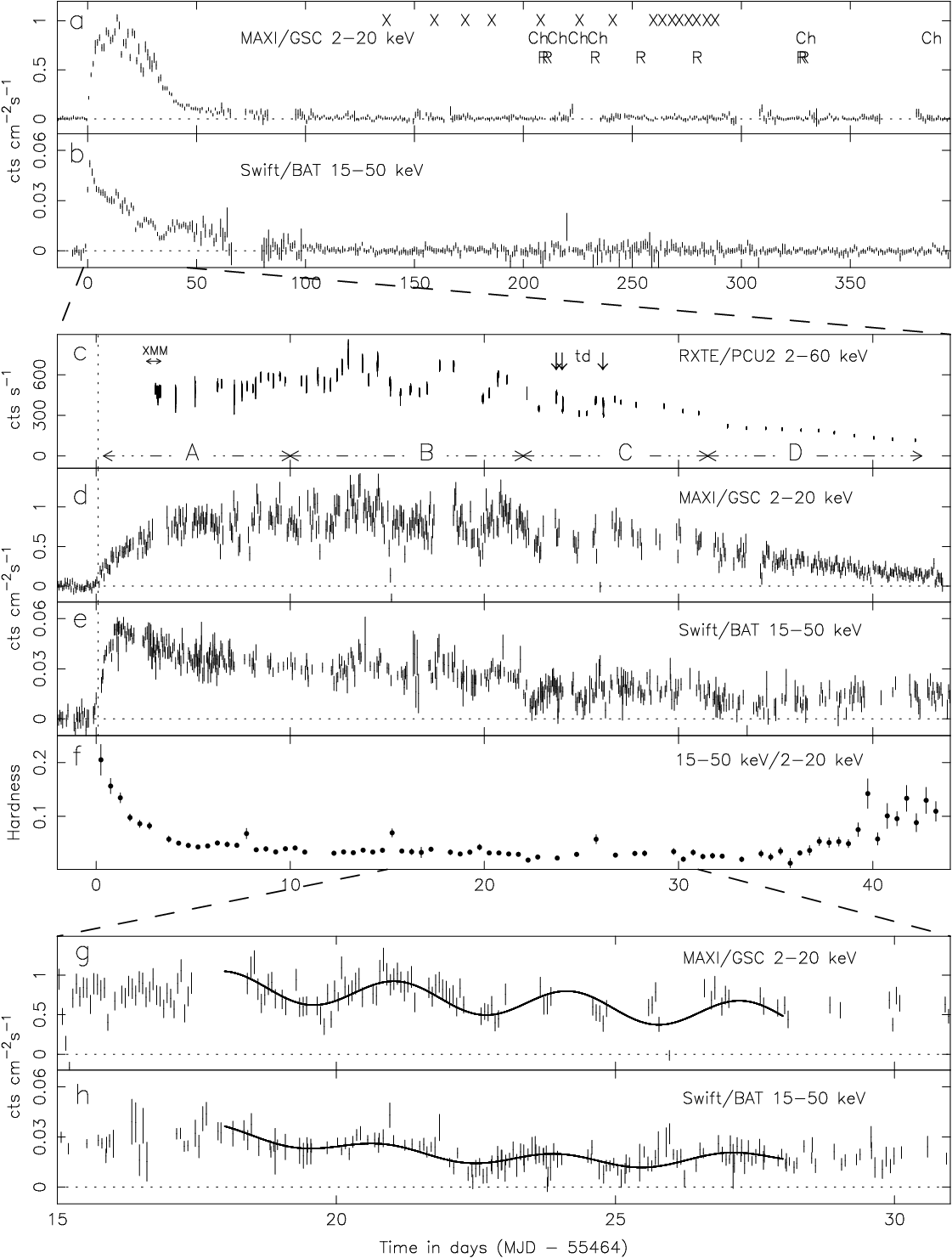}
  \caption{{\it Top panel}: Overview of the 2010/2011 outburst of MAXI\,J1659$-$152. 
({\it a}) MAXI/GSC and ({\it b}) {\it Swift}/BAT daily light curves. 
Note that most of the measurements after day $\sim$100 are non-detections.
In {\it a} we mark the times of the {\it Swift}/XRT (`X'), Chandra (`Ch') and radio (`R') observations taken in 2011 (see Sect.~\ref{overall}). 
{\it Middle panel}: Zoom-in from the top panel, focusing on the main part of the 2010/2011 outburst of MAXI\,J1659$-$152. 
({\it c}) RXTE/PCA PCU2 light curve, at a 16\,s time resolution. 
The data within an observation are connected to guide the eye; errors are not plotted since they are negligible. 
The time-span of the {\it XMM-Newton} observations, discussed in this paper, is indicated in the top-left (`XMM'). 
The times of the transition dips (`td'; see text) are marked by arrows. 
The time spans for the four main epochs described in the text (`A', `B', `C', `D') are indicated at the bottom of this graph.
({\it d}) MAXI/GSC and ({\it e}) {\it Swift}/BAT light curves, showing averages per satellite orbit.
The vertical dotted line marks the time of the first MAXI/GSC detection (Negoro et al.\ 2010).
({\it f}) Hardness values as a function of time. Hardness is defined
as the ratio of the 0.5~day averaged count rates in the {\it Swift}/BAT 15--50\,keV band to the MAXI/GSC 2--20\,keV band. 
{\it Bottom panel}: Zoom-in from the middle panel, focusing on a possible 3-day variation discussed in Sect.~\ref{overall}.
({\it g}) MAXI/GSC and ({\it h}) {\it Swift}/BAT light curves showing the averages over a satellite orbit, between days 15 and 31.
We also show the results of a sinusoidal fit to describe the possible $\sim$3~day variation, plus a constant,
linear and quadratic term to account for the longer-term trend, between days 18 and 28.
The horizontal dotted lines in {\it a, b, d, e, g} and {\it h} correspond to the zero level. 
}
\label{figure1}
\end{figure*}

\subsection{RXTE}
\label{RXTE}

RXTE monitored the source more than once daily throughout the outburst (see Kalamkar et al.\ 2011,
Mu\~noz-Darias et al.\ 2011b, Shaposhnikov et al.\ 2012, Yamaoka et al.\ 2012).
We used 64 PCA (Jahoda et al.\ 2006) observations from 2010 September 28 to November 8 
(see Table~1); the total good-time exposure was about 138\,ks.

For the PCA data we used the {\sc FTOOLS} analysis suite (version 6.9).
We produced light curves from PCU2 with 16-s bins over the full PCA energy range (using Standard 1 data), 
i.e., 2--60\,keV, and over three energy ranges, i.e., 2.1--4.9\,keV, 4.9--9.8\,keV and 9.8--19.8\,keV (using Standard 2 data).
We applied the standard selection criteria for bright sources in our analysis.
We included data when the elevation angle of the source above the Earth horizon was more than 10$^{\circ}$, and used
only stable pointings, i.e., those with offset angles less than 0.02$^{\circ}$. The light curves were corrected for background,
as estimated from the background model for bright sources. 

\subsection{Swift}

We utilised the methods described by Evans et al.\ (2009) to extract XRT (Hill et al.\ 2004, Burrows et al.\ 2005) 
X-ray light curves in the energy range 0.3--10\,keV, with corrections for the effects of pile-up, hot-columns and 
hot-pixels applied.
These light curves, at 100\,s time resolution, were extracted using the most accurate available localisation in the XRT coordinate 
system, derived from late-time PC mode data taken on 2011 February 6, 134 days after the initial detection of 
MAXI\,J1659$-$154, when the source was not affected by pile-up (Kennea et al.\ 2011). 
We used the XRT/WT light curves between 2010 September 25 and October 22 (see Table~1), whenever the pointing offset
was smaller than about 3.5\arcmin. This led to a total good-time exposure of about 123\,ksec.

We used the BAT (Barthelmy et al.\ 2005) 15--50\,keV light curves (see Krimm et al.\ 2006)\footnote{{\tt \tiny http://swift.gsfc.nasa.gov/docs/swift/results/transients/\ index.html}}
generated on 2011 October 25. 
We did not use time bins with less than 500\,s exposure time (see Kennea et al.\ 2011).

\subsection{MAXI}

The Gas Slit Camera (GSC; Mihara et al.\ 2011) is part of the payload of the MAXI mission, 
onboard the International Space Station (ISS). 
Only those data (version 0.3)
when the instrument was operating at a high voltage of 1650\,V were taken into account. 
In this paper we only focus on the light curves averaged over an ISS orbit and on a daily basis, in the total 2--20\,keV band.\footnote{{\tt \tiny http://maxi.riken.jp}}

\section{Results}
\label{results}

The main outburst light curve of MAXI\,J1659$-$152 has already been shown in various papers, as well as a multitude of 
energy range and time resolution combinations. Authors also used different combinations of instruments. 
We refer to Kalamkar et al.\ (2011), Mu\~noz-Darias et al.\ (2011b) and Shaposhnikov et al.\ (2012) 
for the RXTE/PCA light curves of the averages per observation, in various X-ray bands. Yamaoka et al.\ (2012) 
showed the RXTE/PCA data at a time resolution of 16\,s in various X-ray bands.
Kalamkar et al.\ (2011) and Yamaoka et al.\ (2012) also show the one-day averaged MAXI/GSC light curves.
Additionally, the latter authors provided the daily-averaged {\it Swift}/BAT data. Kennea et al.\ (2011) showed
the {\it Swift}/BAT data on a satellite orbit time scale, together with the {\it Swift}/XRT rates at a 100\,s time resolution.
Both Kalamkar et al.\ (2011) and Shaposhnikov et al.\ (2012) provided daily hardness averages based on the 
RXTE/PCA data, whilst Kennea et al.\ (2011) provided hardness curves during the outburst using the {\it Swift}/XRT data.

We here present an overview of the outburst behaviour, combining the information at soft and hard energies, as well as at
various time resolutions.
We first describe the overall outburst light curve and then focus on two independent intensity variations seen, 
i.e., `absorption dips' and `transition dips'. We subsequently present a timing study of the data taken during the 
time period when absorption dips were seen. We use days since MJD\,55464 to describe the epoch of time. 
This date is close to the MAXI/GSC and {\it Swift}/BAT triggers of the outburst (MJD\,55464.10 and 55464.34, respectively, 
see Sect.~\ref{intro}).

\subsection{Overall outburst light curve}
\label{overall}

In the top panel of Fig.~\ref{figure1} we show the overall outburst light curve of MAXI\,J1659$-$152, using daily averages, 
at relatively soft energies (2--20\,keV; Fig.~\ref{figure1}a) and hard energies (15--50\,keV; Fig.~\ref{figure1}b).
After a fast rise of a couple of days, MAXI\,J1659$-$152's soft intensity fluctuates by 20--30\%\ on a daily basis (see also Kalamkar et al.\ 2011), 
on top of a general slow decline in intensity ($\sim$0.1\,cts\,cm$^{-2}$\,s$^{-1}$ per 10 days), up to about day 30. 
It then shows an exponential-like decay (with a decay constant of 
of $\sim$7~days) up to about day 100. In the hard energy band (Fig.~\ref{figure1}b), MAXI\,J1659$-$152 reaches a 
peak in intensity within 3 days after the start. It subsequently decreases in a somewhat irregular fashion 
(but smoother than with respect to the soft energy band), until about day 65 (see also Kennea et al.\ 2011). 
After days 100 and 65, MAXI\,J1659$-$152 falls below the detection limits of MAXI/GSC and {\it Swift}/BAT, respectively.
In Fig.~\ref{figure1}a we have also indicated the times of the post main-outburst X-ray observations with {\it Swift} and 
Chandra, as well as radio observations, as reported in the literature (Kennea et al.\ 2011, 
Yang et al.\ 2011a, 2011b, Yang \&\ Wijnands 2011a, 2011b, 2011c, 2011d, Miller-Jones et al.\ 2011, Jonker et al.\ 2012).

\begin{figure}[top]
\centering
  \includegraphics[height=.31\textheight,angle=-90]{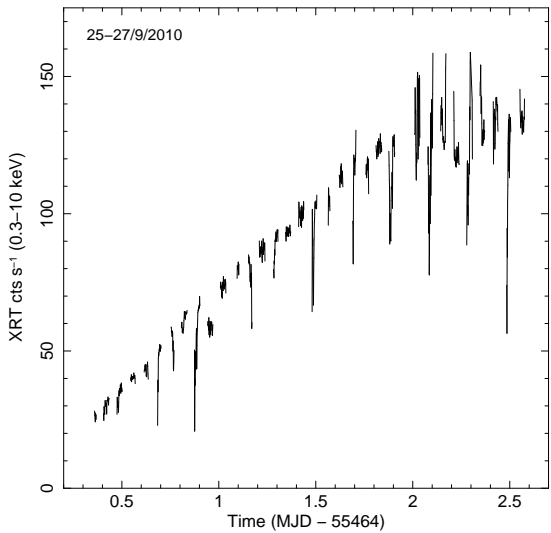}
  \caption{
{\it Swift}/XRT 0.3--10\,keV light curve during the first few days of the outburst of MAXI\,J1659$-$152, 
when {\it Swift} observed the source during every satellite orbit (days 0.3--2.6). 
Dipping activity is clearly apparent on a regular basis.
}
\label{xrt_start}
\end{figure}

In the middle panel of Fig.~\ref{figure1} we zoom in on the main part of the outburst, from just before the start 
of the outburst up to the part when MAXI\,J1659$-$152 was too close to the Sun to be observed during dedicated pointed observations with RXTE and {\it Swift}.

Fig.~\ref{figure1}c shows the outburst as seen by the RXTE/PCA in its whole sensitive energy band (2--60\,keV), 
at a 16\,s time resolution. The variability during an observation in the first part of the main outburst and up to before the 
soft X-ray peak of the main outburst (i.e., up to about day 10, which we denote outburst epoch A) is mainly due to periodic dips in the light curves, which will be 
described in more detail in the next subsection, Sect.~\ref{absorptiondips}.
From before the peak of the outburst up to about day 22 (outburst epoch B), variability within an observation is still seen, albeit less strong.
This variability is due to flaring during the observations (see, e.g., Fig.~\ref{RXTE_lc_hardness_dip}c).
During outburst epoch B the average flux varies between 15--20\%\ from observation to observation
(see also Kalamkar et al.\ 2011). 
On days 22--31.5 (outburst epoch C), apart from a general decreasing trend, the intensity appears to be varying between a low and a high value. 
Flux variations on a similar scale are seen within three observations of that period (days 23.7, 24.0 and 26.1, 
see Sect.~\ref{transitiondips}). The last part of the main outburst (days 30.5--44.1; outburst epoch D) shows a rather smooth,
but not linear or power-law/exponential like (see below), decay. 
We note that the observations during outburst epoch D have only a relatively short duration (see Table 1); they exhibit no strong variability like that seen in the earlier outburst epochs.

\begin{figure*}[top]
\sidecaption
\includegraphics[width=12cm]{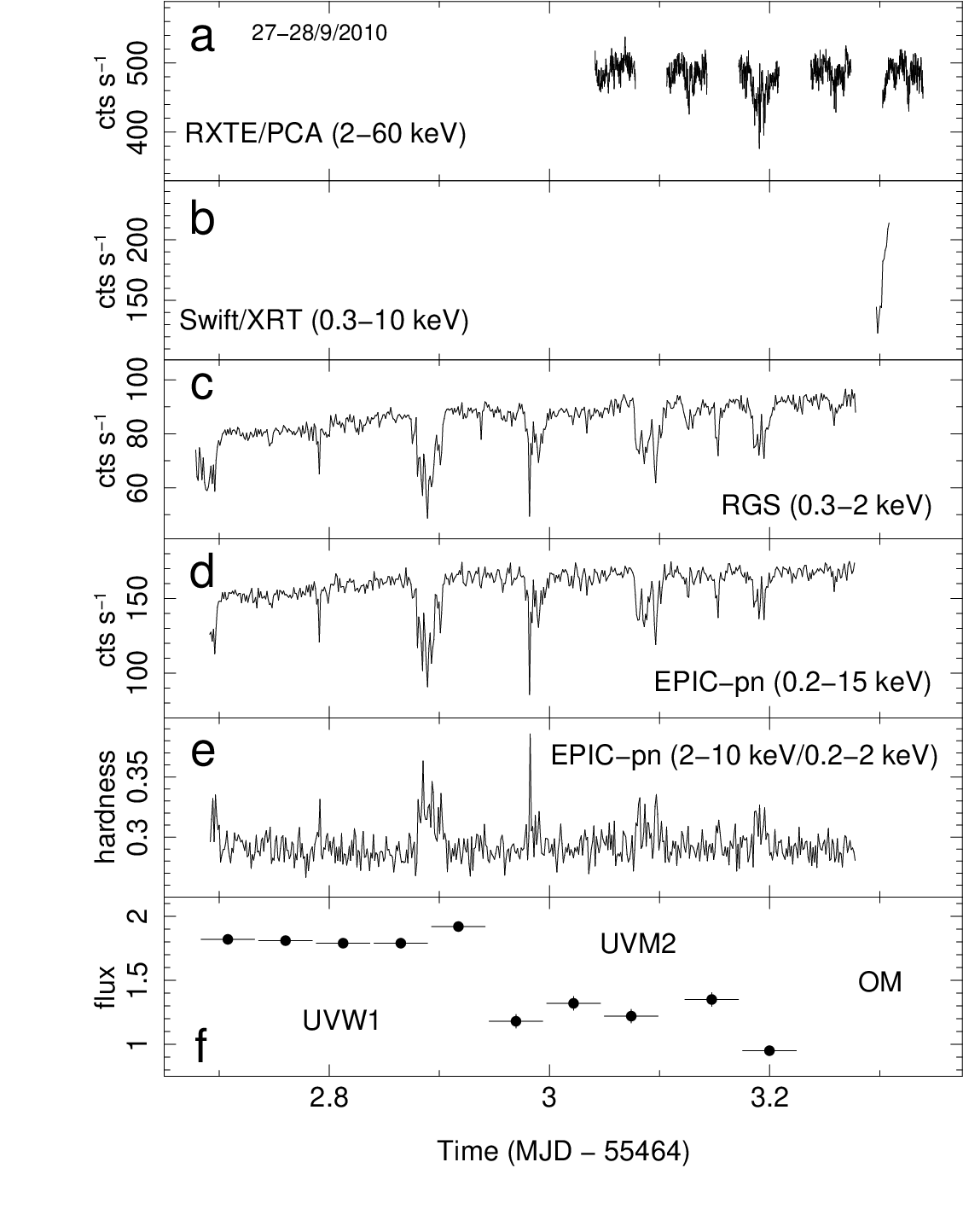}
\caption{
RXTE/PCA ({\it a}), {\it Swift}/XRT ({\it b}), {\it XMM-Newton}/RGS ({\it c}),
{\it XMM-Newton}/EPIC-pn ({\it d}), and EPIC-pn hardness (ratio of count rates in the 2--10\,keV to 0.2--2\,keV
bands; {\it e}) and {\it XMM-Newton}/OM (UVW1 and UVM2; {\it f}) curves on days 2.7 to 3.3.
The time resolution is 16\,s for the RXTE/PCA data and 100\,s for the data from the other X-ray instruments. 
The X-ray data points are connected for clarity. The OM flux is in units of
10$^{-15}$\,erg\,cm$^{-2}$\,s$^{-1}$\,\AA$^{-1}$, see also Table~2. Note that UVW1 and UVM2 do not cover the same
wavelength range (see Sect.~\ref{XMM}).
\vspace{1cm}
}
\label{lightcurves}
\end{figure*}

In Figs.~\ref{figure1}d and e the MAXI/GSC and {\it Swift}/BAT light curves of the outburst are presented, 
integrated over a satellite orbit.
We note that the {\it Swift}/XRT (and {\it Swift}/UVOT, see Kennea et al.\ 2011) coverage was up to day 27.3 
(see Table~1 and Kennea et al.\ 2011).
The difference in rise time to maximum between the soft and hard energy band, noted above, can be clearly seen.
This is also borne out by the hardness curve shown in Fig.~\ref{figure1}f: just after discovery, the outburst is hard, and then
softens during the following week (see also Kennea et al.\ 2011, using {\it Swift}/XRT). Near the end of the main outburst, 
i.e., around day 35, MAXI\,J1659$-$152's radiation hardens again (see also Kalamkar et al.\ 2011, Shaposhnikov et al.\ 2012, using RXTE/PCA). 
The end of the main outburst is well covered by the MAXI/GSC (see also above), 
and, as noted earlier, it shows a smooth, power-law/exponential like decay in the 2--20\,keV band.
The hardening around day 35 and the fact that RXTE/PCA is sensitive to hard ($\gtrsim$20\,keV) X-rays causes the RXTE/PCA light curve to deviate from the power-law/exponential like decay described at softer X-rays ($\lesssim$20\,keV).

Between about days 18 and 28, the MAXI/GSC and {\it Swift}/BAT intensities seem to modulate on a several day time scale
(Figs.~\ref{figure1}d and e). Before and after this time period this is not evident.
A closer look at the MAXI/GSC and {\it Swift}/BAT light curves (Fig.~\ref{figure1}g and h) shows 
a possible periodic variation on an about 3-day time scale, for about 3 cycles. 
Indeed, a pure linear trend does not describe the data well in this time period: $\chi^2_{\rm red}=3.3$ for 81 degrees of freedom (dof) 
and $\chi^2_{\rm red}=2.1$ for 123 dof, for the MAXI/GSC and {\it Swift}/BAT light curves, respectively.
Adding a quadratic term does not significantly improve the situation ($\chi^2_{\rm red}=3.4$ for 80 dof and 
1.8 for 122 dof, respectively). A sinusoidal plus constant, linear and quadratic term does significantly improve the fit, 
although it is still not ideal ($\chi^2_{\rm red}=2.2$ for 77 dof and $\chi^2_{\rm red}=1.3$ for 119 dof, respectively).
The modulation may thus be not purely sinusoidal. Assuming {the signal is real, we derive} a period of 3.15$\pm$0.05~days 
(uncertainty is determined by using $\Delta\chi^2=1$) in the {\it Swift}/XRT light curve and 3.04$\pm$0.09~days in the 
{\it Swift}/BAT, i.e., consistent with each other. 
The phase of the sinusoid is about 0.25~days earlier for the {\it Swift}/BAT with respect to that of the MAXI/GSC.

\subsection{Recurrent intensity variations}

\subsubsection{Absorption dips}
\label{absorptiondips}

Periodic drops of intensity, or dips, are observed shortly after the start of the outburst, at day 0.3, up to day 8.2, 
first in the {\it Swift}/XRT light curves (e.g., Fig.~\ref{xrt_start}, days 0.3--2.6, Fig.~\ref{lightcurves}b, day 3.3), 
then in the {\it XMM-Newton}/RGS and EPIC-pn light curves 
(Fig.~\ref{lightcurves}c and d, days 2.7--3.3), and finally in the RXTE/PCA curves 
(e.g., Fig.~\ref{lightcurves}a, days 3.0--3.4 and Fig.~\ref{RXTE_lc_hardness_dip}a, day 7.1). 
A clear recurrence time of $\simeq$0.1 days is best observed in the {\it XMM-Newton}/RGS and EPIC-pn light curves, thanks to the continuous coverage. 
In addition to the periodic dips, a linear rise in the out-of-dip intensity is observed in the {\it Swift}/XRT light curve (Fig.~\ref{xrt_start}; see also Kennea et al.\ 2011), 
which extends throughout the {\it XMM-Newton} observations (Fig.~\ref{lightcurves}c and d) and is consistent with the MAXI/GSC light curve 
(Fig.~\ref{figure1}d), where the source reached its first plateau at soft X-rays, $\lesssim$20 keV, around day 4 (i.e., just after the {\it XMM-Newton} observations).

The dips show irregular structure which lasts between about 5 and 40\,min. Occasionally,
the {\it XMM-Newton}/RGS and EPIC-pn light curves show shallower dip activity at half the recurrence time 
(see, e.g., Figs.~\ref{lightcurves}c and d near day 3.15). The depth of the dips varies between $\simeq$90\%\ and 50\%\ of 
the average out-of-dip-interval intensity in the light curves of {\it Swift}/XRT and {\it XMM-Newton}/RGS and EPIC-pn 
(Figs.~\ref{xrt_start} and \ref{lightcurves}), extracted with a time resolution of 100\,s. 

\begin{figure}[top]
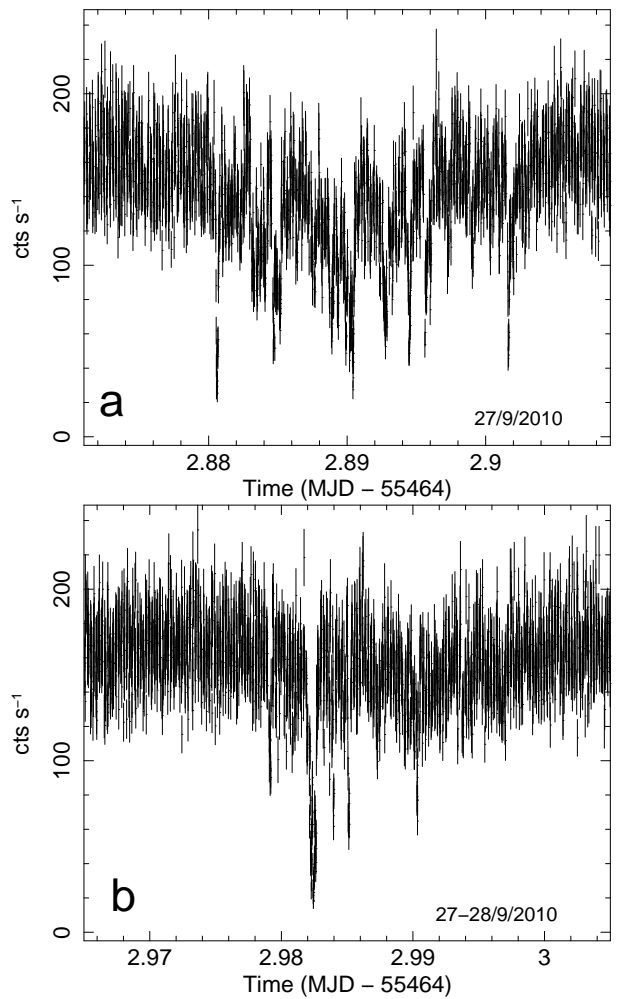

\centering
  \includegraphics[height=.34\textheight,angle=-90]{dip1.eps}
  \includegraphics[height=.34\textheight,angle=-90]{dip2.eps}
  \caption{
Zoom in from Fig.~\ref{lightcurves}d on the {\it XMM-Newton}/EPIC-pn (0.2--15\,keV) light curves around the time of dip 
activity around day 2.9 ({\it a}) and 3.0 ({\it b}), respectively. The time resolution is 1\,s. 
All data with a fractional exposure of more than 0.5 per bin are shown.
}
\label{dips}
\end{figure}

\begin{figure*}[top]
\centering
  \includegraphics[height=.5\textheight,angle=-90]{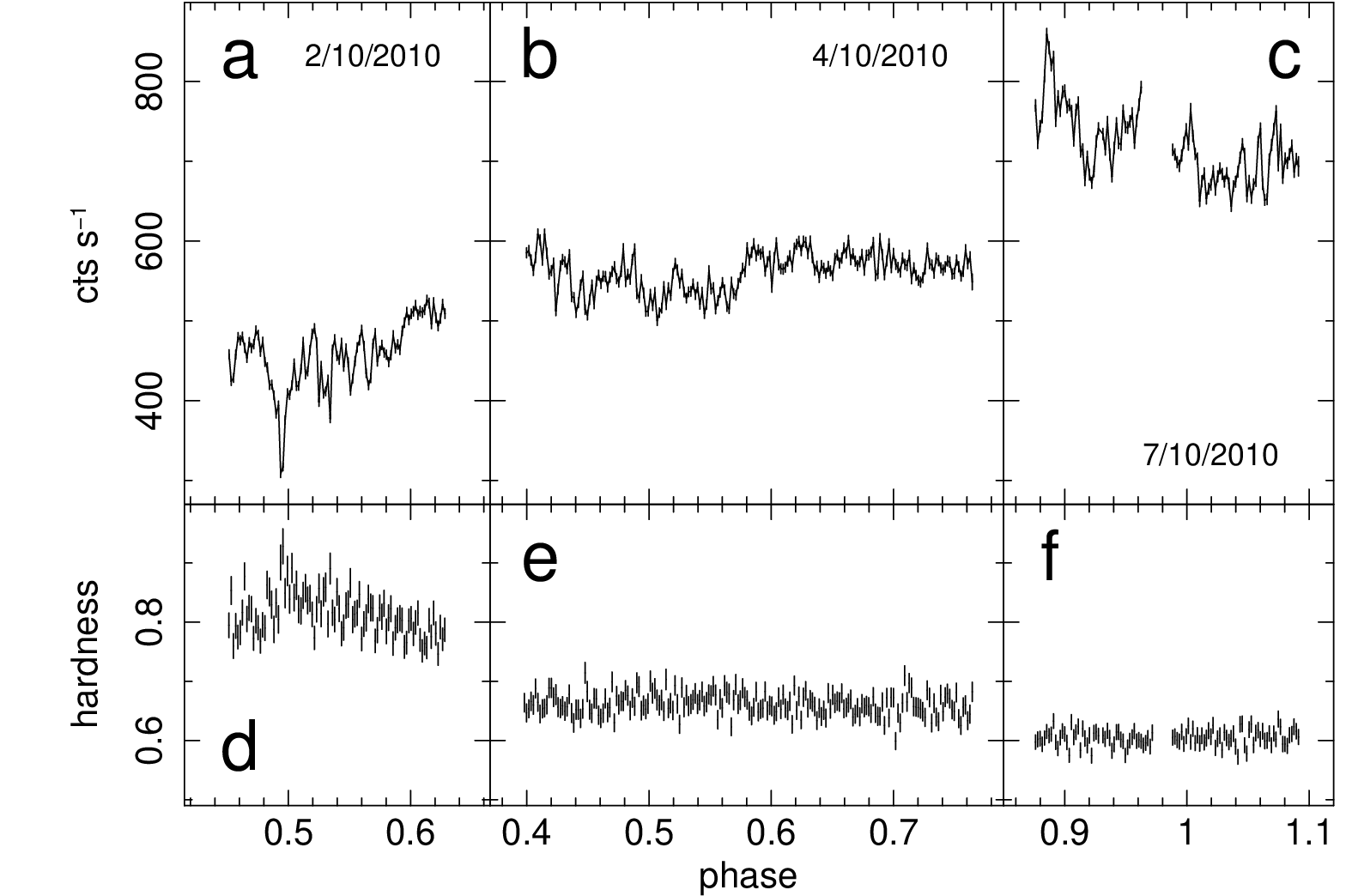}
 $^{~~~~}$  \includegraphics[height=.46\textheight,angle=-90]{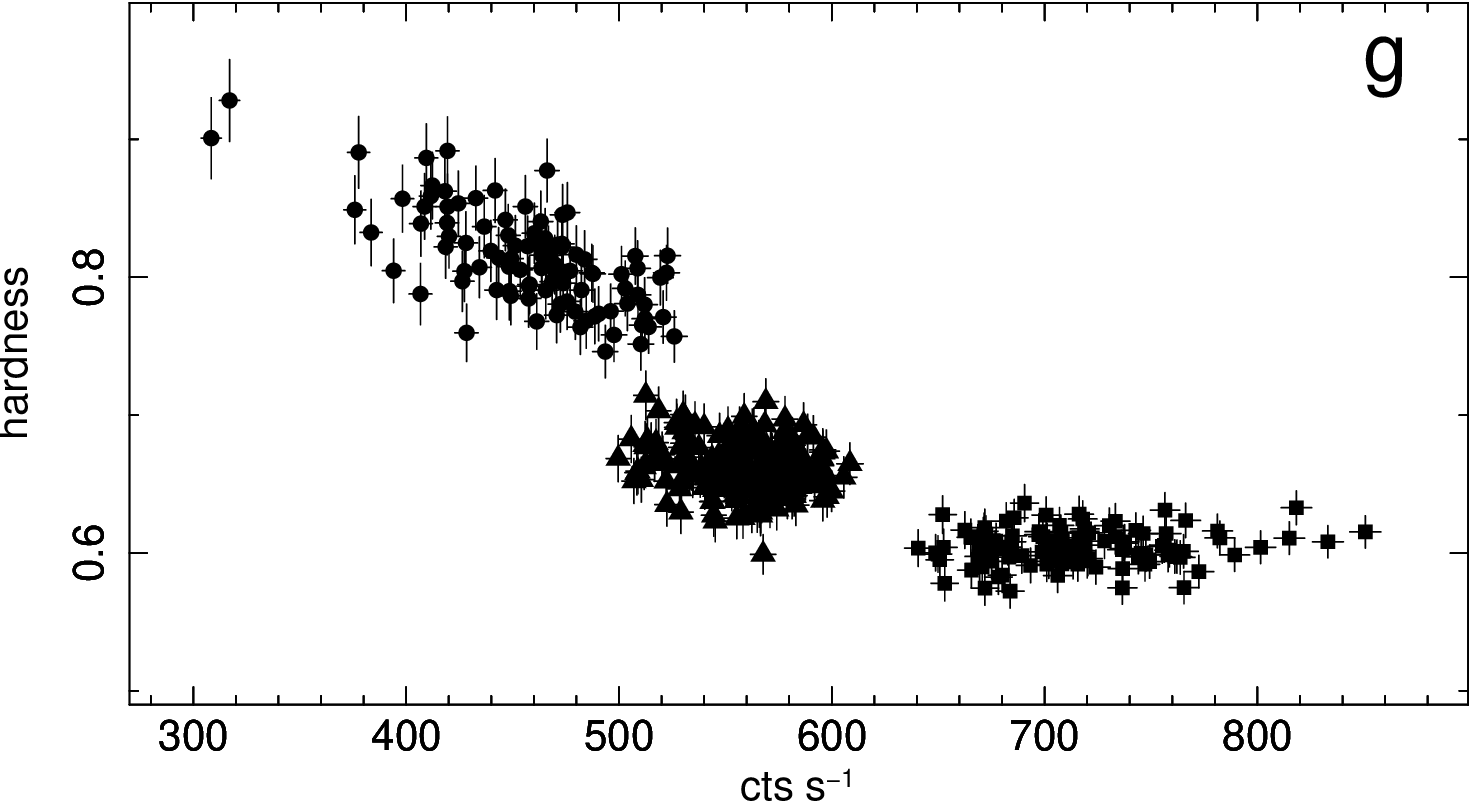}
  \caption{Intensity ({\it top}) and hardness ratio ({\it middle}) curves during three RXTE/PCA observations 
(day 7.1: {\it a}, {\it d}; day 9.1: {\it b}, {\it e}; day 12.7: {\it c}, {\it f}), using PCU2 data at a time resolution of 16\,s.
The light curves are for the 2--60\,keV band, whilst the hardness is defined as the ratio of the count rates
in the 4.9--9.8\,keV and 2.1--4.9\,keV bands. The curves are folded on the absorption dipping activity ephemeris
(see Sect.~\ref{timing}). {\it Bottom}: hardness versus intensity diagram (HID) for the same three observations.
Day 7.1: filled circles; day 9.1: filled triangles; day 12.7: filled squares). 
}
\label{RXTE_lc_hardness_dip}
\end{figure*}

In Fig.~\ref{dips} we show two epochs of the dip activity at a higher
time resolution of 1\,s as seen by the {\it XMM-Newton}/EPIC-pn. The two epochs
are one dip-cycle apart, i.e., about 0.1\,day. Clearly, the dip morphology
changes from cycle to cycle. Fast dipping activity is
observed, which can last up to 30\,min. These fast dips have a
duration of $\lesssim$30\,s, and often as small as 1\,s. Occasionally, before and/or after the fast dips
the intensity reaches the persistent level seen outside the dipping
intervals. During the fast dips the intensity can drop down to about 15\%\ of the persistent
intensity, indicating that the shallower dips observed in Fig.~\ref{lightcurves} are
a consequence of averaging a smaller number of fast dips in coarser time bins.

Next, we examine the hardness values of the {\it XMM-Newton}/EPIC-pn (Fig.~\ref{lightcurves}e) and RXTE/PCA (see Fig.~\ref{RXTE_lc_hardness_dip}) 
light curves. The hardness ratio is defined as the ratio of the count rates in the 
2--10\,keV to the 0.6--2\,keV bands for the EPIC-pn data, and 4.9--9.8\,keV to the 2.1--4.9\,keV bands for the PCA data.
During the dips the source hardens. This colour behaviour during dips is
also confirmed by the {\it Swift}/XRT data (see Kennea et al.\ 2011).
We observe in particular a stronger hardening as the dip becomes deeper. 
As the out-of-dip intensity increases, the dipping becomes 
shallower and the changes in the hardness ratio less pronounced (see Figs.~\ref{RXTE_lc_hardness_dip}a and b). 
This is shown in more detail in Fig.~\ref{RXTE_lc_hardness_dip}g, 
where we plot the changes of hardness ratio as a function of count rate. 
The shape of the curve traced by the PCA data from day 7.1 to 9.1 to 12.7 is remarkably similar to 
the shape of the curve shown in figure~4 of Boirin et al.\ (2005) for the classical LMXB dipper 4U\,1323$-$62 as the source moves 
from deep dipping to shallow dipping and finally to a persistent state.\footnote{Note 
that the differences in the numbers are due to the different instruments used for the curves, 
RXTE/PCA in Fig.~\ref{RXTE_lc_hardness_dip}g and {\it XMM-Newton}/EPIC-pn in figure~4 of Boirin et al.\ (2005).}
Using the absorption dip activity ephemeris from Sect.~\ref{timing}, we find that the dips recur at phase 0.4--0.6 (see, e.g., Fig.~\ref{RXTE_lc_hardness_dip}a--f).

Since the dip morphology and hardness behaviour resembles the dip phenomenon encountered in various high-inclination
X-ray binaries (Sect.~\ref{D_absorptiondips}), we refer to these dips as `absorption dips'.

We note that UV data simultaneous to the {\it XMM-Newton}/EPIC-pn data are available from the {\it XMM-Newton}/OM 
(see Fig.~\ref{lightcurves}f). Although the UV is variable, there does not
seem to be a correlation with X-ray intensity, in particular with
the X-ray dipping. Unfortunately, however, the time resolution
is too coarse to investigate in detail the UV light curve on the
X-ray dip structure time scale. Therefore, we do not discuss the UV data any further.

\subsubsection{Transition dips}
\label{transitiondips}

\begin{figure*}[top]
\centering
  \includegraphics[height=.5\textheight,angle=-90]{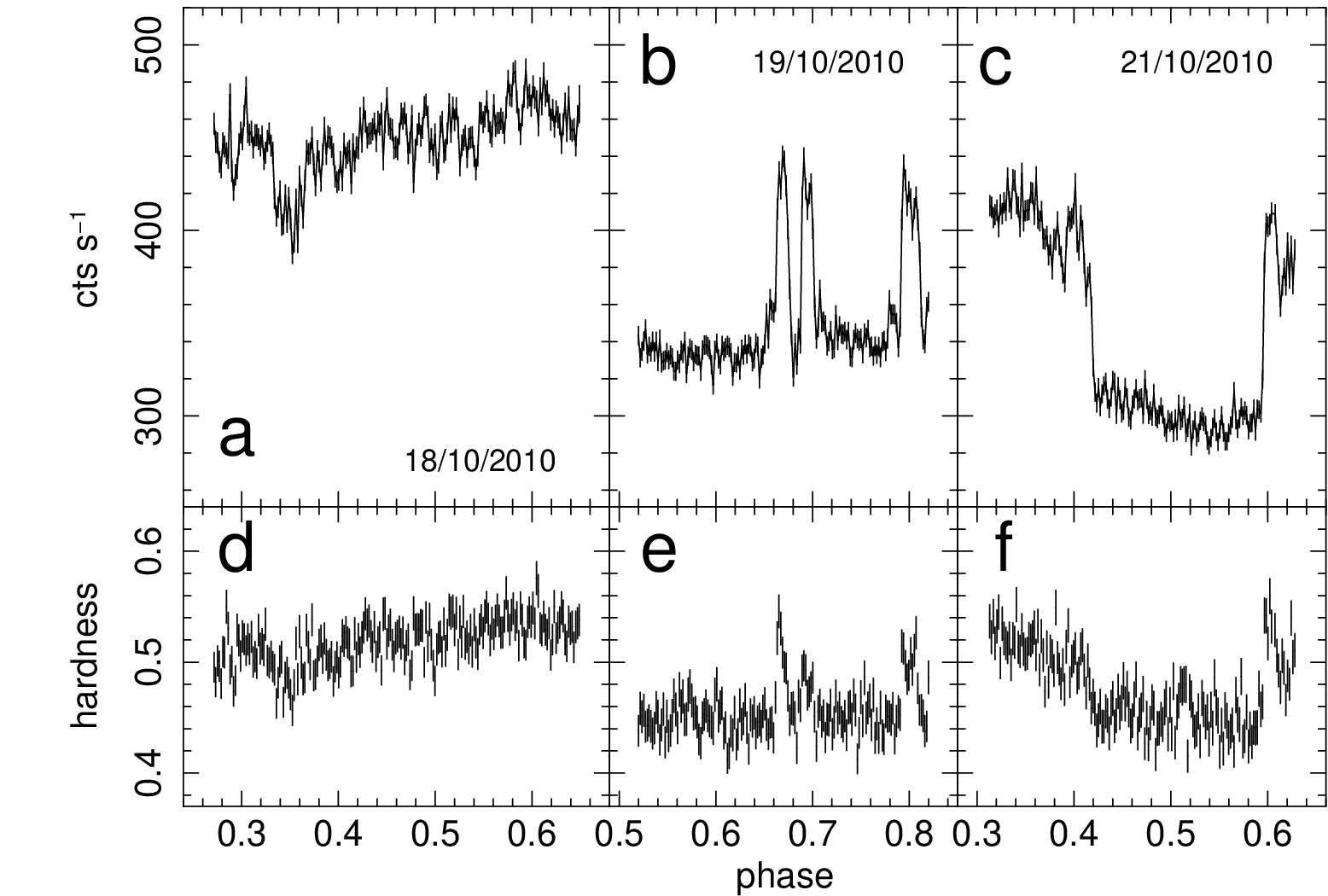}
 $^{~~~~~~~~}$  \includegraphics[height=.48\textheight,angle=-90]{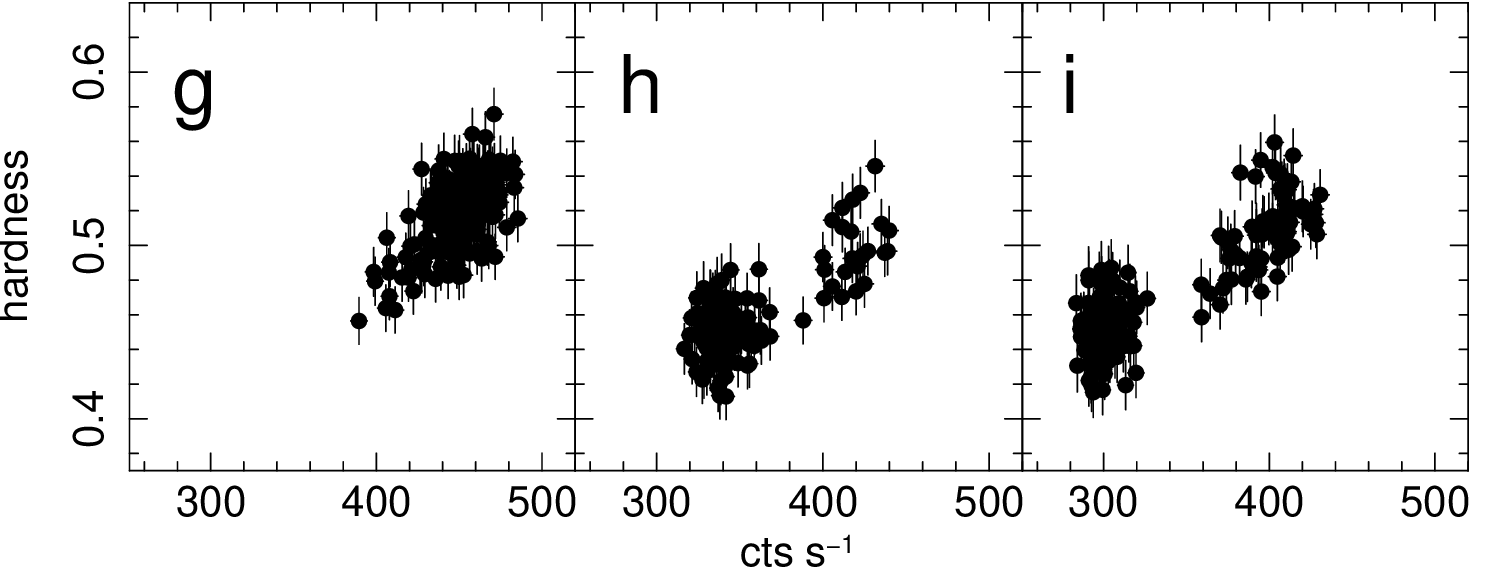}
  \caption{Intensity ({\it top}) and hardness ratio ({\it middle}) curves during the RXTE/PCA observations with 
transition dips (day 23.7: {\it a}, {\it d}, respectively; 24.0: {\it b}, {\it e}, respectively; day 26.1: {\it c}, {\it f}, respectively), 
using PCU2 data at a time resolution of 16\,s.
The light curve is for the 2--60\,keV band, whilst the hardness is defined as the ratio of the count rates 
in the 4.9--9.8\,keV and 2.1--4.9\,keV bands. The curves are folded on the absorption dipping activity ephemeris 
(see Sect.~\ref{results}). {\it Bottom}: hardness versus intensity diagram (HID) for the same three observations as above
({\it g}, {\it h}, and {\it i} for day 23.7, 24.0 and 26.1, respectively), using the same data. 
}
\label{RXTE_lc_hardness_tr}
\end{figure*}

On three occasions (days 23.7, 24.0 and 26.1; see Fig.~\ref{RXTE_lc_hardness_tr}), the X-ray light curves
show further pronounced variations. 
The RXTE/PCA light curve on day 23.7 (Fig.~\ref{RXTE_lc_hardness_tr}a) resembles that seen during observations
with absorption dips. However, the hardness (Fig.~\ref{RXTE_lc_hardness_tr}d, g) correlates with the
X-ray intensity, in contrast to the absorption dips.
The RXTE/PCA light curves on days 24.0 and 26.1 (Fig.~\ref{RXTE_lc_hardness_tr}b and c) show the presence of two intensity levels 
which differ by about 30\%, between which the source fluctuates. The transitions are fast, i.e., 
they occur within tenths of seconds. 
The time spent in the upper level is about 1--2\,min on day 24.0 and at least 6--14\,min on day 26.1. The time spent 
in the lower level is about 12 to at least 20\,min on day 24.0 and about 26\,min on day 26.1.
These light curves have been referred to as 'flip-flop' light curves (see Kalamkar et al.\ 2011). However, the intensity
fluctuations correspond to fast source state transitions (see Sect.~\ref{transitiondips_discussion}), so we refer to them as `transition dips'.
The intensity variations seen during the transition dips are more or less of the same order as 
the variations in the average intensity from observation to observation in outburst epoch C
(see Fig.~\ref{figure1} and Sect.~\ref{overall}; see also Kalamkar et al.\ 2011).
Contrary to Kalamkar et al.\ (2011), we find clear hardness changes during the transition dips light curves (Fig.~\ref{RXTE_lc_hardness_tr}e and f).
The hardness behaviour is similar to that seen on day 23.7, again opposite to that seen during the absorption dip episodes, 
i.e., the source is harder at the upper intensity level, and softens when the source transits to the lower level 
(Figs.~\ref{RXTE_lc_hardness_tr}h and i). 

Although the phasing\footnote{By folding the data on the recurrence time of the absorption dip activity, 
$\simeq$0.1~day (see Sects.~\ref{absorptiondips} and \ref{timing}).} of the transition dips on day 26.1 
is consistent with the phasing of the absorption dips 
(see Sect.~\ref{absorptiondips}), the phasing of the transition dips on days 23.7 and 24.0 is clearly not.
Observations in between days 24.0 and 26.1 do not show any dipping behaviour (even between phase 0.4 and 0.6), and 
indicate that the transition dips do not recur with the 0.1~day time scale.

\subsection{Timing analysis of absorption dip activity}
\label{timing}

The {\it Swift}/XRT, {\it XMM-Newton}/EPIC-pn and RGS, and RXTE/PCA data indicate that the absorption dips occur regularly, 
i.e., about every $\sim$0.1~day (Sect.~\ref{absorptiondips}), from the start of the main outburst up to about day 10 (outburst epoch A, see Sect.~\ref{overall}). 
In LMXB dippers (Sect.~\ref{D_absorptiondips}) absorption dips recur with the orbital period.
Using our rather large time base line we can establish an accurate period of the recurring dip activity in MAXI\,J1659$-$152. 

We performed a Lomb-Scargle (LS; Lomb 1976, Scargle 1982) period search, as well as a
Phase Dispersion Minimisation (PDM; Stellingwerf 1978) period search, on our data sets taken during outburst epoch A. 
The datasets of each instrument were treated separately. 
The LS and PDM searches were done over the period range 0.01--0.5\,days, with a frequency interval of 0.001\,day$^{-1}$.
For the PDM search we used 20 phase bins with a phase bin width of 
0.05\footnote{The choice of number of phase bins and phase bin width is 
rather arbitrary; tests with various numbers of phase bins and different phase bin width yielded consistent results.}.
In the LS periodogram a peak indicates a dominating period in the data set, whilst in 
a PDM periodogram a minimum indicates a dominating period.

The error on a period found was computed by constructing 1000 synthesised data sets.
These data sets were obtained by distributing each data point around
its observed value, by an amount given by its error bar multiplied by a number output by a Gaussian
random-number generator with zero mean and unit variance. 
The measured standard deviation of the positions of the deepest minima in $\Theta$ or highest peak in power
in the resulting periodograms was taken as the error.
These latter periodograms were done in a narrow range around $\simeq$0.1~day, i.e., 
between 0.0909 and 0.1111 days, with a frequency interval of 0.00009~days.

To check the significance of the peaks and minima found in the LS and PDM diagrams, respectively, 
we randomised the data and calculated the resulting periodogram. 
The randomisation was done by keeping the time tags and randomly distributing the intensities of the data sets which were input to the periodogram programs.
This was repeated 1000 times and the resulting averaged periodogram was used to evaluate the significance of the 
peaks/minima. We find that the values of the power (LS) or amplitude $\Theta$ (PDM) are narrowly distributed around 1 for all periods investigated: the variances are 1 for the LS values of all the instruments, and 0.001, 0.001, 0.002 and 0.007 for the PDM values of the RGS, EPIC-pn, XRT and PCA, respectively. 
We refer to this level as the noise level.

Another way to characterise the uncertainty in the period, which is more conservative, is to use the width of the peak or minimum 
of the periodograms, e.g., the half-width at half maximum or minimum (HWHM). 

The results for the different instruments are discussed in the next subsections, and a summary of the best-found 
periods near 0.1~days with their associated errors is given in Table~\ref{periodsearch}.
We find, that the HWHM values are a factor of 25--200 times larger than those derived by 
the measured standard deviations of the positions of the peaks or minima, as described above. 
Since the spread in the best-found periods is of the order of the HWHM values, we use these values as
a final indicator of the uncertainty in the derived periods.

\begin{figure*}[top]
\centering
  \includegraphics[height=.34\textheight,angle=-90]{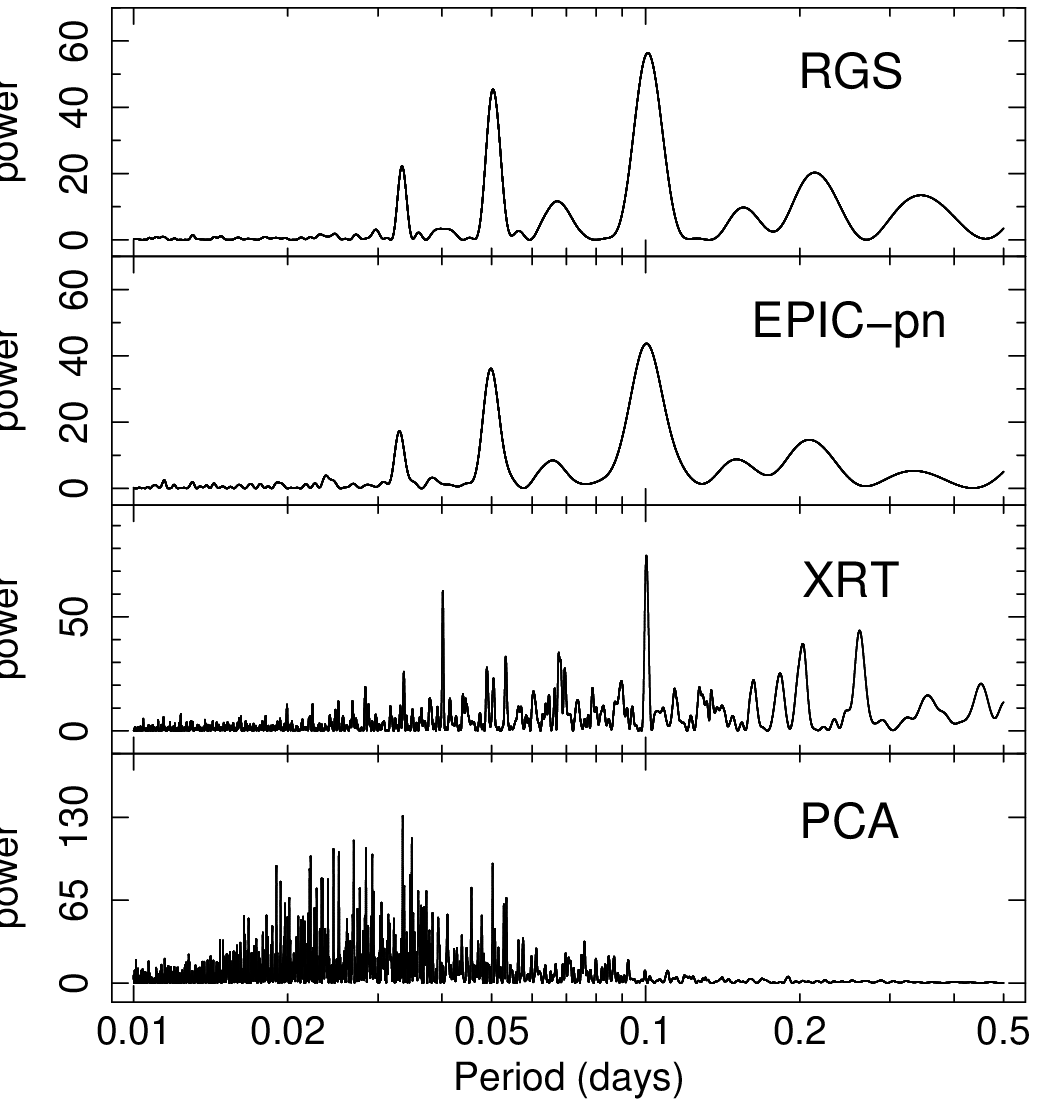}
  \includegraphics[height=.34\textheight,angle=-90]{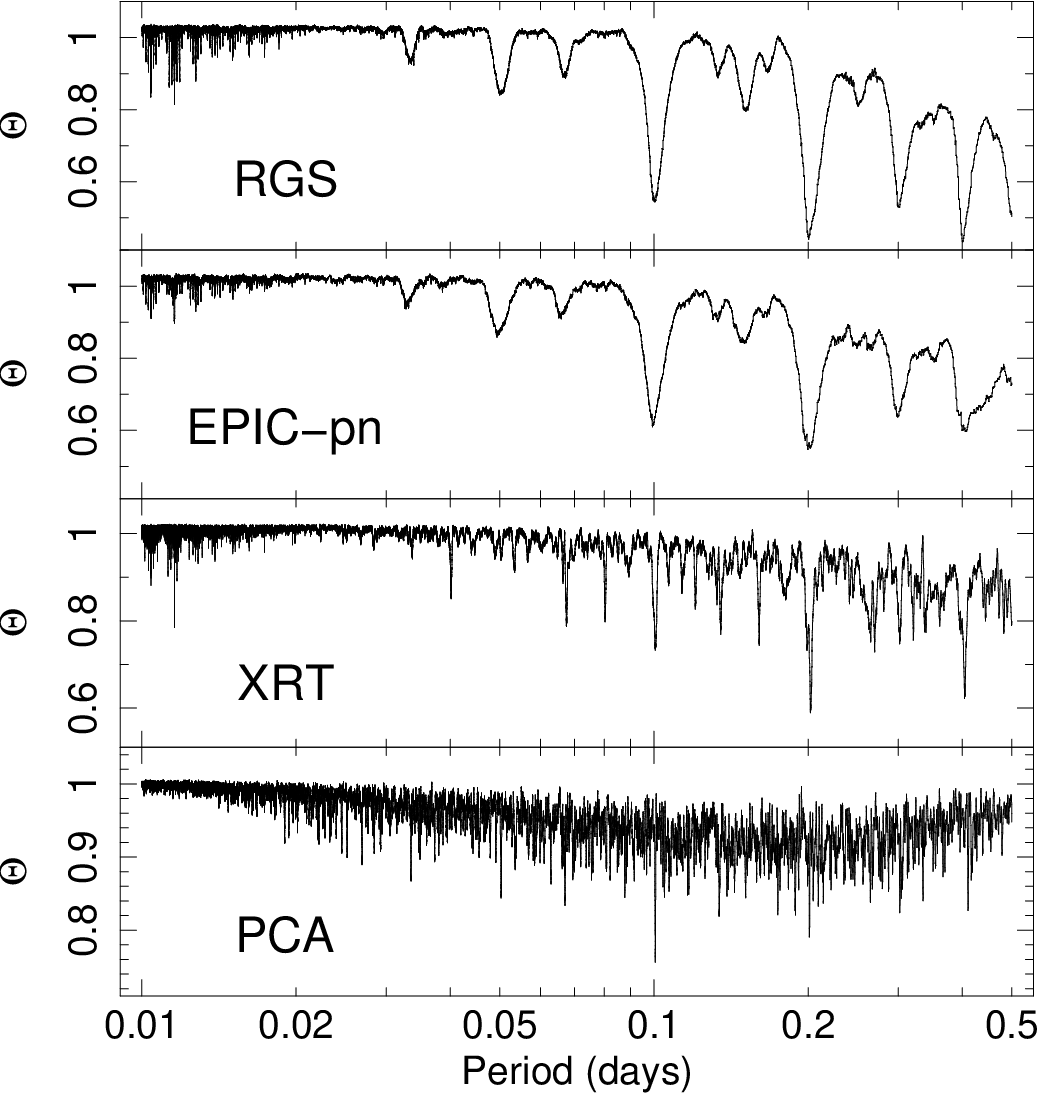}
  \caption{
Results of the Lomb-Scargle ({\it left}) and Phase Dispersion Minimisation ({\it right}) period analyses (see text) of the {\it XMM-Newton}/EPIC-pn ({\it top}), {\it XMM-Newton}/RGS ({\it second panel}), {\it Swift}/XRT ({\it third panel}) and RXTE/PCA ({\it bottom}) datasets. 
}
\label{pdm_scargle}
\end{figure*}

\subsubsection{XMM-Newton/EPIC-pn and RGS}
\label{xmmsearch}

We used the EPIC-pn and the RGS data with 100\,s time resolution. 
The highest peak in the LS periodograms is at a period near 0.1~day (top two left panels of Fig.~\ref{pdm_scargle}), 
i.e., 0.1004$\pm$0.0090\,days and 0.1010$\pm$0.0071\,days, respectively, for the EPIC-pn and RGS data sets. 
Three deep minima are visible in the RGS and EPIC-pn PDM periodograms, near 0.1, 0.2 and 0.4~days (top two right panels of Fig.~\ref{pdm_scargle}). 
The deepest minimum is not at the best period found with the LS search, however.

Inspection of the folded light curves on the three periods found with the PDM, reveals that only when folding the data on the 
$\simeq$0.1~day period is the absorption dip activity (i.e., at count rates $\lesssim$140\,c\,s$^{-1}$ and $\lesssim$75\,c\,s$^{-1}$ 
for the EPIC-pn and RGS, respectively) clustered within 0.2 in phase space. For the other two periods,
the same absorption dip activity is distributed along all phases. 
This strengthens our conclusion that the fundamental period is near 0.1~days, since absorption dip activity is expected to
occur at restricted orbital phases (see Sect.~\ref{D_absorptiondips}).

\subsubsection{Swift/XRT}
\label{xrtsearch}

A LS and PDM search on the XRT data during outburst epoch A did not reveal any significant period, except for the satellite 
orbital period around the Earth (see below). This we attribute to the variation 
in intensity of the overall main outburst light curve, which is of the same order as the drops in intensity during the 
absorption dips (see Fig.~\ref{xrt_start} and Kennea et al.\ 2011). 
The increase in the out-of-dip intensity is rather gradual during the first 6 days. Thereafter, the out-of-dip light curve
varies irregularly on a time scale of days. We, therefore, first detrended the data using 
a multi-order polynomial (see also Kennea et al.\ 2011); a third-order polynomial describes the overall 
out-of-dip light curve up to about outburst day 6 sufficiently well.
Including data after day 6 contaminated our period search significantly, and had the effect
of diminishing the peak and minima in our LS and PDM searches, respectively, near 0.1~day.
We, therefore, use these data only up to day 6, in the remainder of this subsection.

The highest peak in the LS periodogram is at 0.1003$\pm$0.0010~days (see Fig.~\ref{pdm_scargle}, left panel).\footnote{We note 
that the period reported by Kennea et al.\ (2011), 0.1008$\pm$0.0037~days, was determined using the LS search on 
the detrended first 12.8~days of the XRT/WT data. Their error in the period 
is derived by fitting a Gaussian to the Lomb-Scargle periodogram around the peak, and taking a value of 2.7$\sigma$, where 
$\sigma$ is the width of the Gaussian.}
Several minima are found in the PDM periodogram (Fig.~\ref{pdm_scargle}, right panel), including those seen at the same periods as 
in the PDM periodograms of the {\it XMM-Newton} data (Sect.~\ref{xmmsearch}). The minimum at 0.1005$\pm$0.0010~days is 
clearly not the dominant period in this PDM search. However, inspection of the folded light curves on the various periods 
with peaks or minima in the LS and PDM searches, respectively, show that, again, only for the period near 0.1~day, the dip 
activity is clustered within 0.2 in phase space, as expected for absorption dips (see Sect.~\ref{D_absorptiondips}).

We note that performing the period search after renormalising the XRT data in a similar manner as we did for the 
RXTE data (see Sect.~\ref{pcasearch}) did not reveal 
significant power at the above reported periods. 
This is because the XRT observations were in general shorter
than the RXTE/PCA, resulting in averages which are higher when there is a dipping period, and therefore
the variations due to dipping are diminished.

\begin{figure}[top]
\centering
  \includegraphics[height=.34\textheight,angle=-90]{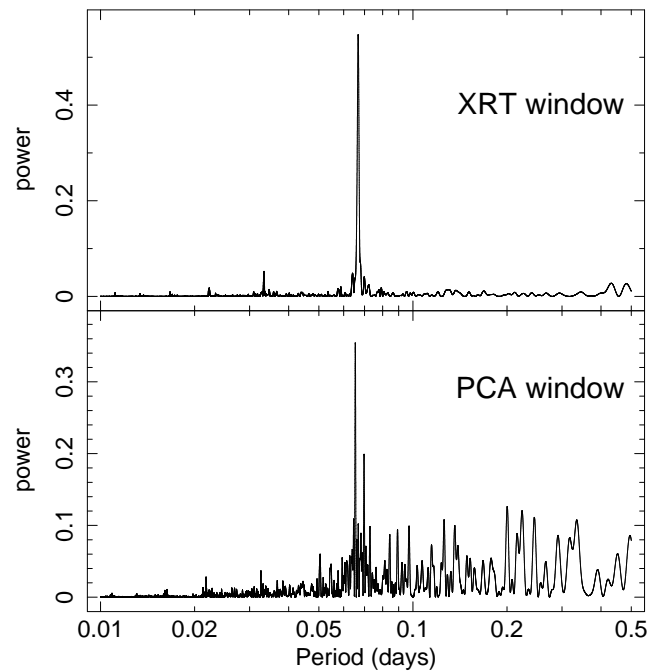}
  \caption{
The Fourier transforms of the window functions using the RXTE/PCA ({\it top}) and {\it Swift}/XRT ({\it bottom}) data.
The peaks in power density spectra are consistent with the satellite orbital periods around the Earth, i.e., 
0.065~days (94\,min) and 0.067~days (96\,min), respectively for the RXTE/PCA and {\it Swift}/XRT.
}
\label{window}
\end{figure}

To inspect whether the fundamental period is related to the data sampling (such as due to the satellite orbit), 
we constructed a Fourier transform (FT) of the window function. This window function was determined by setting the 
intensities of the time series to zero. For the FT we used the same frequency settings as the LS and PDM searches. 
The largest peak in the FT periodogram is evidently at the satellite orbit period around the Earth of about 0.067~days
(Fig.~\ref{window}, top). Thus our best period is not related to any peak in the FT periodogram, and can thus be 
considered to be intrinsic to the source (see also Kennea et al.\ 2011).

\subsubsection{RXTE/PCA}
\label{pcasearch}

The overall variability of the PCA data (see Sect~\ref{overall} and Fig.~\ref{figure1}c),
prevented the LS and PDM periodograms to show any significant peaks or minima, respectively, except at the satellite
orbital period around the Earth (see below). 
Detrending the PCA data as done for the XRT did not improve our period searches, however. 
This is due to the fact that the PCA data sampling (as well as the energy range) is different from that of the XRT data. 
Moreover, there are significant day-to-day variations in the average out-of-dip PCA intensity, which cannot 
be described by a simple polynomial. We, therefore, renormalised the PCA light curves during the whole 
outburst epoch A in our search for periodicities. For each observation interval 
(generally corresponding to an RXTE satellite orbit) we determined the mean count rate. This value was subtracted 
from the light curves corresponding to each of these observation intervals.

The lowest minimum in the PDM periodogram is at a period of 0.10058$\pm$0.00022 days (Fig.~\ref{pdm_scargle}, bottom right). 
A PDM search in three energy bands (2--5\,keV, 5--10\,keV and 10--20\,keV, see Sect.~\ref{RXTE}) shows the deepest minima at the 
same period (see Table~\ref{periodsearch}).
Again, the period quoted above does not coincide with the strong peak at about 0.065~days in the FT of the PCA window function (Fig.~\ref{window}, bottom panel). 

A LS search on the PCA data did not reveal a peak near $\simeq$0.1\,day period, nor near any of the other periods found 
in the {\it XMM-Newton} and XRT period searches (Fig.~\ref{pdm_scargle}, bottom left).\footnote{The highest peak in the LS periodogram
is at 0.0335~days. However, folding the data at this period reveals the absorption dips to be distributed along
all phases.}
We attribute this to the highly non-sinusoidal nature of the modulation in the 
renormalised light curves at 16\,s time resolution, and the possible imperfection of our method of renormalisation for the LS search. 

\subsubsection{Recurrence period of dipping activity}
\label{ephemeris}

\begin{table*}
\caption{Results of the PDM and LS period searches in the light curves for the different instruments.}
\begin{tabular}{lllcccc}
\hline
\multicolumn{1}{l}{} &
\multicolumn{3}{c}{PDM} &
\multicolumn{3}{c}{LS} \\
\multicolumn{1}{l}{Instrument} &
\multicolumn{1}{c}{period} &
\multicolumn{1}{c}{error} &
\multicolumn{1}{c}{HWHM} &
\multicolumn{1}{c}{period} &
\multicolumn{1}{c}{error} &
\multicolumn{1}{c}{HWHM} \\
\multicolumn{1}{l}{(energy)} &
\multicolumn{1}{c}{(days)} &
\multicolumn{1}{c}{(days)} &
\multicolumn{1}{c}{(days)} &
\multicolumn{1}{c}{(days)} &
\multicolumn{1}{c}{(days)} &
\multicolumn{1}{c}{(days)} \\
\hline
RGS (0.3--2\,keV)      & 0.10070 & 0.00013 & 0.0042 & 0.100985 & 0.000098 & 0.0071 \\
EPIC-pn (0.2--15\,keV) & 0.09931 & 0.00006 & 0.0049 & 0.100394 & 0.000190 & 0.0090 \\
XRT (0.3--10\,keV)     & 0.10054 & 0.00004 & 0.0010 & 0.100282 & 0.000011 & 0.0010 \\
PCA (2--60\,keV)       & 0.100582 & 0.000001 & 0.00022 &  ---   &   ---   &   --- \\
PCA (2--5\,keV)        & 0.100590 & 0.000018 & 0.00021 &  ---   &   ---   &   --- \\
PCA (5--10\,keV)       & 0.100590 & 0.000019 & 0.00023 &  ---   &   ---   &   --- \\
PCA (10--20\,keV)      & 0.100580 & 0.000024 & 0.00022 &  ---   &   ---   &   --- \\
\hline
\end{tabular}
\label{periodsearch}
\end{table*}

The {\it XMM-Newton} and XRT LS period searches and the PCA PDM period search reveal one common best period, i.e., 
$\sim$0.1~day, with peaks/minima in the periodograms well above/below the noise level.
Inspection of the individual light curves of the various instruments, and the folding of the data on the 
$\simeq$0.1~day period leading to restricted phase range of dipping (see below), supports the main period to be at that value.
We attribute the peaks/minima at half this period in the {\it XMM-Newton} periodograms to the intermediate dipping episodes seen.
The peaks/minima at multiple times the above quoted period are due to the fact that the morphology of the dipping changes from cycle to cycle, as well as the fact that not all dipping periods are sampled well enough.

The RXTE/PCA provides the strongest constraint on the period.
We, therefore, conclude that the fundamental period of dipping activity is at 0.10058$\pm$0.00022~days.
We arbitrarily set the zero point of the absorption dip activity ephemeris to the first data point of the first 
RXTE/PCA observation at day 3.0. Therefore, the absorption dip activity ephemeris is 
$T_0 = {\rm MJD}\,55467.039561 + 0.10058(22)\times E$, where E is the cycle number of the period.

\begin{figure}[top]
\centering
  \includegraphics[height=.35\textheight,angle=-90]{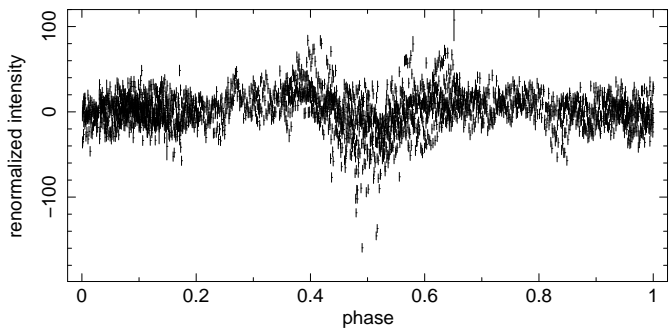}
  \caption{
The detrended RXTE/PCA (2--60\,keV) 16\,s data folded on the ephemeris given in Sect.~\ref{ephemeris}.
}
\label{folded}
\end{figure}

The folded detrended PCA curve during the dip epoch using this ephemeris is shown in Fig.~\ref{folded}.
The dipping activity has a duty cycle of about 0.2 in phase, around phase 0.5. 
The mid-point between start and end of the duty cycle corresponds to T$_{0,\rm dip}= {\rm MJD}\,55467.0904\pm 0.0005$.

\section{Discussion}
\label{discussion}

\subsection{Origin of intensity variations}

\subsubsection{Absorption dips}
\label{D_absorptiondips}

We find regular absorption dips in the X-ray light curves during the outburst of MAXI\,J1659$-$152 between outburst days 0.3 and 8.2. 
Absorption dips recur at the orbital period of the system and are thought to be caused by obscuration by material located in 
a thickened outer region (`bulge') of the accretion disk due to its interaction with 
the inflowing gas stream from the companion (e.g., White \&\ Swank 1982, Walter et al.\ 1982; see discussion below). 
The presence of absorption dips allows a precise measurement of the orbital period and is a signature of high inclination
(e.g., White \&\ Swank 1982, White \&\ Mason 1985; see also D\'\i az Trigo et al.\ 2009). 

We determine the recurrence period of the dips to be 2.414$\pm$0.005~hrs (see also Kuulkers et al.\ 2012a, Kennea et al.\ 2011). 
By analogy with other classical dippers, we identify this period with the orbital period of a system.
The fastest revolving binary was
Swift\,J1753.5$-$0127 (3.2443\,hrs; Zurita et al.\ 2008, Durant et al.\ 2009). 
As suggested by \mbox{Kuulkers} et al.\ (2010d, 2012a), if the compact object in
MAXI\,J1659$-$152 is indeed a black hole (Kalamkar et al.\ 2011, Kennea et al.\ 2011, Mu\~noz-Darias et al.\ 2011b, Shaposhnikov et al.\ 2012, Yamaoka et al.\ 2012), 
its $\simeq$2.4\,hrs period is the shortest among the currently known BHXB sample (see, e.g., Ritter \&\ Kolb 2003).\footnote{A possible exception may be 
Swift\,J1357.2$-$0933, based on an indirect estimate of the orbital period of 2\,hr by Casares et al.\ (2011).
We note, however, that the outburst amplitude in the optical ($\sim$6~mag, Rau et al.\ 2011) is incompatible with such a short period, based
on the empirical relation between the outburst amplitude and orbital period (Shabaz \&\ Kuulkers 1998), see also Sect.~\ref{class}.
Another short-period binary candidate, MAXI\,J1305$-$704, was put forward recently by Kennea et al.\ (2012b). This new transient (Sato et al.\ 2012)
showed absorption dips. A periodicity of $\sim$1.5\,hrs was reported (Kennea et al.\ 2012b), but this was put into doubt by Kuulkers et al.\ (2012b).
It has been suggested that MAXI\,J1305$-$704 is a BHXB based on 
outburst optical amplitude, blue optical spectral energy distribution and hard X-ray spectrum (Greiner et al.\ 2012; see also Kennea et al.\ 2012a), as well as 
the occurrence of a state transition (Suwa et al.\ 2012). However, these are features which are seen in neutron star LMXBs as well (see Suwa et al.\ 2012, Kennea et al.\ 2012b).}

We constrain the inclination of MAXI\,J1659$-$152 to be between about 65$^{\circ}$ and 80$^{\circ}$ from the presence of the periodic absorption dips,
due to material in the line of sight that obscures up to about 90\%\ of the total emission at given cycles, and the absence of eclipses. 
We base the lower limit on the size of the bulge (and not on the disk opening angle which has been generally estimated to be $\simeq$12$^{\circ}$, e.g., de Jong et al.\ 1996, Bayless et al.\ 2010).
White \&\ Holt (1982) estimated the size of the bulge responsible for absorption dips as 19$^{\circ}$$\pm$6$^{\circ}$ for the LMXB 4U\,1822$-$37. 
Taking this as typical for LMXBs, we can thus set a lower limit on the inclination of 65$^{\circ}$. 
We note that, if the accretion disk is tilted or warped, the lower limit
for the inclination could be as low as 55$^{\circ}$, taking into account that for generic LMXBs, a disk tilt of about 10$^{\circ}$
is expected (Foulkes et al.\ 2010).
An upper limit for the inclination of 80$^{\circ}$
is derived from the absence of eclipses (e.g., Horne 1985), the spectral type of the companion star (M5V, see Sect.~\ref{unevolved}), and the fact that 
the companion is filling its Roche lobe (see, e.g., Motch et al.\ 1987). 
Here we are assuming that the source of emission is point-like (for an extended source one would not be able to see full eclipses for any of the LMXBs).

The absorption dips in MAXI\,J1659$-$152 share many of the properties of classical absorption dipping systems. 
They change from period to period, they are fast, and the obscuration can be large, i.e., down to about 90\%\ of the total intensity
(see, e.g., White \&\ Mason 1985, Parmar \&\ White 1988). 
Boirin et al.\ (2005) and D\'\i az Trigo et al.\ (2006) were able to model the changes in both the narrow X-ray absorption features
and the continuum during the dips from all the bright dipping LMXBs observed by {\it XMM-Newton} by an increase in the column density and a 
decrease in the amount of ionisation of a photo-ionised absorbing plasma. The changes in the hardness ratio observed in the dips in 
MAXI\,J1659$-$152 are consistent with absorption by neutral and photo-ionised plasma, in the sense that they
are energy dependent. A further support to the existence of neutral and photo-ionised plasma is the presence of various stages of dipping: 
persistent, shallow and deep dipping states. The fact that dips become shallower and less energy dependent as the count rate increases 
could be a consequence of the photo-ionised plasma becoming more and more ionised and transparent as it is illuminated by the X-rays of the central region. 
However, we note that a definite confirmation of an increase of neutral and ionised plasma during dips for MAXI\,J1659$-$152 is only possible after spectral analysis.

Absorption dipping in other LMXBs occurs mainly around orbital phase 0.7--0.9, where eclipses are expected at phase zero if we view the accretion disk edge-on,
i.e., when the companion star is closest to us and in front of the neutron star (e.g., Parmar \&\ White 1988). 
Occasionally, absorption dips with a 0.5 phase difference with respect to the phase at which `regular dips' occur are observed. 
These `anomalous dips', or `secondary absorption dips', were also seen in other dipping systems (such as XB\,1916$-$053, e.g., 
White \&\ Swank 1982, Walter et al.\ 1982, Smale et al.\ 1988, Boirin et al.\ 2004). They are explained as being due to material 
migrating to the other side after impact with the disk, or the accretion stream partly freely overflowing the disk or bouncing of the disk rim and then overflow
(see, e.g., Frank et al.\ 1987, Armitage \&\ Livio 1998, and references therein).
In the latter case the flow may impact the disk near the circularisation radius, either causing a second bulge (see, e.g., Frank et al.\ 1987, Armitage \&\ Livio 1998, 
and references therein) or bouncing of the disk again (Kunze et al.\ 2001). 

The absorption dips appear for only part of the outburst in MAXI\,J1659$-$152. This has been observed already for other BHXBs undergoing an outburst, 
like 4U\,1630$-$47 and GRO\,J1655$-$40 (Kuulkers et al.\ 1998, 2000, Tomsick et al.\ 1998). It is plausible that for transient BHXBs, changes in the 
accretion mode cause the appearance or disappearance of dips. Kuulkers et al.\ (2000) interpreted the (deep) absorption dips during the rise and 
plateau phase of the outburst in GRO\,J1655$-$40 as due to an absorbing medium which is filamentary in nature. These filaments could be due to the stream of
material coming from the companion star splashing into the accretion disk and overflowing above and below the impact region
(e.g., Livio et al.\ 1986; see Kuulkers et al.\ 2000, and references therein). If the inclination is high enough, the impact region itself comes also into the line of 
sight (e.g., Frank et al.\ 1987). 
However, the presence of absorption features all around the orbit for neutron stars (e.g., Parmar et al.\ 2002) shows that at least part of the 
photo-ionised plasma is distributed equatorially along the whole plane of the disk, indicating that absorption is due to a structure in the 
disk rather than by filaments. In that scenario, the cause for the disappearance of dips in BHXBs could be, e.g., a strong ionisation of the plasma 
in bright (but hard)
states of the outburst, which renders the plasma transparent and therefore invisible to us. 
Alternatively, a change of the structure of the disk could diminish the thickness of the bulge and cause the absorption dips to disappear.

\subsubsection{Transition dips}
\label{transitiondips_discussion}

As noted by Kalamkar et al.\ (2011), during the second epoch of the outburst rapid and sharp flux variations --- transition dips --- 
were seen, resembling the `flip-flop' and `dip' light curves in, e.g., GX\,339$-$4 (Miyamoto et al.\ 1991), GS\,1124$-$68 
(Takizawa et al.\ 1997) and XTE\,J1859+226 (Casella et al.\ 2004). In all these cases 10--20\% changes in intensity were seen. 
We find that MAXI\,J1659$-$152 softens when the intensity decreases, in contrast to the hardening seen during the absorption dips.
In the case of GX\,339$-$4 (Miyamoto et al.\ 1991) and XTE\,J1859+226 (Casella et al.\ 2004)
significant hardness changes could be discerned as well, with behaviour similar to that seen for MAXI\,J1659$-$152. The clear 
difference in hardness behaviour of the absorption and transition dips, as well as the fact that they occurred in well separated 
phases of the outburst, suggests that the two phenomena have a different origin. 

\begin{figure*}[top]
\sidecaption
\includegraphics[width=12cm]{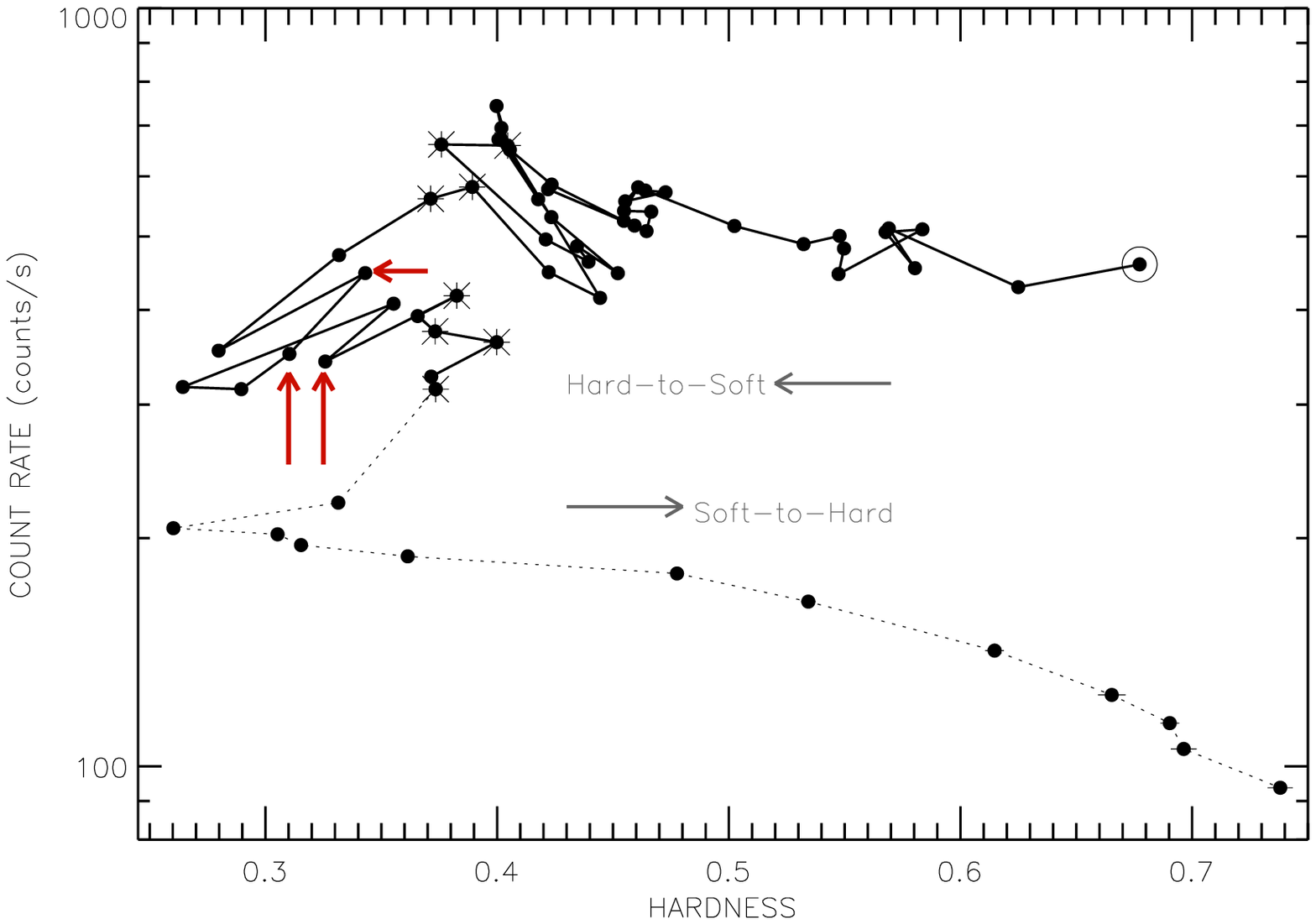}
\caption{
{\it Top:} Hardness-intensity diagram obtained using all the RXTE/PCA observations available (adapted from Mu\~noz-Darias et al.\ (2011b). 
Intensity corresponds to the count rate 
in the 2--15\,keV band and hardness is defined as the ratio of counts in 6.1--10.2\,keV and 3.3--6.1 keV bands. Each point corresponds to the
average over an observation. 
Observations with a star correspond to those with a so-called type-B QPO in the power-density spectra (see text). A solid line joins consecutive 
observations starting from observation with a big, open circle (top right). Observations taken after the last type-B QPO are joined by a dotted line. 
The three arrows mark the observations with clear transition dips on days
23.7, 24.0 and 26.1. All observations between days 0 and 10, i.e., when absorption dips occur, have hardness values $\gtrsim$0.45.
\vspace{1cm}
}
\label{HID}
\end{figure*}

The transition dips in MAXI\,J1659$-$152 occurred during the first soft excursion of the source (see Fig.~\ref{HID}). During this 
period the count rate differences between consecutive observations were of the same order as the count rate changes seen during the 
transition dips (with the softer observations having lower count rates), suggesting that additional transitions took place between 
observations. 

In the other sources in which transitions dips have been seen, the transitions were often accompanied by pronounced changes in the 
power density spectra (Miyamoto et al.\ 1991, Takizawa et al.\ 1997, Homan et al.\ 2001, Casella et al.\ 2004). The power density 
spectra from the two observations that showed the transitions dips in MAXI\,J1659$-$152 (days 24.0 and 26.1) were not of high enough 
quality to detect significant changes in the power-density spectral properties. However, by analysing the averaged power density spectra from the 
combined high and low count rate levels during the first `soft excursion', Kalamkar et al.\ (2011) were able to see indications for 
an additional broad bump around 7--8\,Hz in the power density spectra of the low count rate selection. We note that the transitions in 
other sources often involve so-called `type B' quasi-periodic oscillations, QPOs, either in the low or high count rate level power density spectra. The broad excess seen 
by Kalamkar et al.\ (2011) is too broad to be identified as a type B QPO, and is more likely to be a peaked noise component.

Transition dips are most likely the result from instabilities in the inner accretion flow (see Miyamoto et al.\ 1991 for an 
example interpretation), but their exact origin remains unknown. 
The observations of MAXI\,J1659$-$152 do not provide significant new insights into the nature of these instabilities, but they do 
show that transition dips can also be found in states that are slightly softer than those in which they have been observed in other 
sources (i.e., states in which type B QPOs are observed). This trend is very clearly seen in Fig.~\ref{HID}, 
where the type B QPOs all occur within a hardness range of 0.36--0.40, whilst all transition dips occur at hardness $\lesssim$0.36. 

Around the time of the occurrence of transition dips we find some marginal evidence for the soft and hard X-ray light curves to modulate on a
$\simeq$3~day period. 
We speculate that this period may be related to a disk precession period. Systems with extreme mass ratio's (i.e., $q\lesssim 0.33$), like MAXI\,J1659$-$152,
are vulnerable to a 3:1 orbital resonance within the accretion disk.
This causes the disk to be eccentric and to slowly precess on time scales of days to weeks, which may be discernable in the light curves
(Whitehurst 1988, Whitehurst \&\ King 1991, Lubow 1991a,b; see also Haswell et al.\ 2001). 
Periodic variations are also foreseen in this model with a period slightly longer than the orbital period. However, 
in LMXBs this phenomenon is inclination dependent: in systems with a high orbital inclination only orbital modulations due
to the heated face of the companion star are expected (Haswell et al.\ 2001). 
This is consistent with the fact that the X-ray and optical light curves show the same (orbital) period (see Sect.~\ref{intro}).
The transition dips occurred during the time of the $\simeq$3~day modulation. Possibly, the non-axisymmetric accretion disk
modulates the inner accretion flow, giving rise to the sporadic transition dips in the X-ray light curves.

\subsection{Binary system and evolutionary state}
\label{binsys}

Using various existing empirical relationships one can get 
an idea of the dimensions of the system. To derive uncertainties on the resulting values, where appropriate we
randomly distributed the observed values and equation parameters around their values using a Gaussian distribution with
width equal to their errors, and assumed that all values and parameters are independent.
The spread in the resulting values was used as the uncertainty.
We note that parts of the discussion in the following subsections already appeared in 
Kuulkers et al.\ (2012a) and Kennea et al.\ (2011). We here update various of the values, based on our more refined analysis. 
We indicate where Kennea et al.\ (2011) derived comparable properties.

\subsubsection{Mass and radius of the companion star}
\label{mandr}

\paragraph{4.2.1.1 Unevolved main sequence companions}
\label{unevolved}
~\\
~\\
We use here only the observed empirical mass-radius relationships for donor stars in Cataclysmic Variable (CV) binaries:
(i) Using the linear relationship for CV secondaries by Smith and Dillon (1998; their equations 9 and 12) we derive 
the mass of the companion, $M_2 = 0.19$\,M$_{\odot}$, and its radius, $R_2 = 0.24$\,R$_{\odot}$, with an estimated uncertainty 
of 0.05\,M$_{\odot}$ and 0.02\,R$_{\odot}$, respectively.
(ii) Using the fact that the orbital period of $\simeq$2.4\,hr of MAXI\,J1659$-$152 puts it right in the 
so-called "period gap" of CV-binaries, the empirical mass-radius relation for CV secondaries of Knigge et al.\ (2011; their figure 4) 
yields a very precise mass estimate of $M_2 = 0.20$\,M$_{\odot}$, and a radius of $R_2 = 0.26$\,R$_{\odot}$ with a rather large 
uncertainty of 0.04\,R$_{\odot}$.  
Since both estimates are fully consistent with each other, and since the mass estimate from Knigge et al.'s relation is very precise, 
we will adopt the latter values as the real parameters of the companion star in the system. The mass and radius correspond to spectral 
type M5V if the star would be on the main sequence. However, it is well known that secondaries in CVs and LMXBs are often not precisely 
on the main sequence, and tend to be a bit overluminous for their masses.
The fact that the companion here is still filling its Roche lobe (as it is transferring mass), whilst its orbital period is located in 
the period-gap of CVs implies that the companion is not a normal main sequence star, since in that case it would at that period not be 
filling its Roche lobe. It must therefore be a nuclearly somewhat evolved star (see below).

\paragraph{4.2.1.2 Nuclearly evolved stripped companions}
\label{stripped}
~\\
~\\
If the companion star was originally much more massive and became a nuclearly evolved star, it will now be a stripped evolved star. 
Such a star will be He-rich and will obey a different mass-radius relation. We argue here that the companion is indeed such a 
star and started out its life with a mass between 1.0 and 1.5\,M$_{\odot}$, and underwent considerable nuclear evolution. 
The reasons for this are as follows. Several authors, starting with Pylyser \&\ Savonije (1988), made calculations of the 
evolution of LMXBs driven by the internal evolution of the companion star in combination with orbital angular 
momentum loss by `magnetic braking' and gravitational waves (see Pfahl et al.\ 2003, for recent evolutionary calculations for LMXBs). 
This evolution leads to such stripped evolved companion stars. 
The reason why the companion of MAXI\,J1659$-$152 must be such an evolved star is that its orbital period of $\sim$2.4\,hr falls 
right in the middle of the `period gap' of CVs. In CVs with a normal H-rich main-sequence companion the 
mass transfer is driven solely by orbital angular momentum loss, due to magnetic braking and gravitation radiation
(see, e.g., Howell et al.\ 2001, Knigge et al.\ 2011). 
Such systems evolve to shorter and shorter orbital periods; when they arrive at the upper edge of the period gap, at 
$P_{\rm orb} \simeq 3$\,hr, they stop transferring matter, because magnetic braking stops 
(Spruit \&\ Ritter 1983, Howell et al.\ 2001, Knigge et al.\ 2011). The companion star which was 
somewhat out of thermal equilibrium then shrinks, the orbit also shrinks --- though slower --- due to gravitational wave losses. 
The mass transfer resumes only when the system has reached the lower edge of the period gap at about $P_{\rm orb} \simeq 2$\,hr. This type of 
evolution will hold for all binaries with a low-mass main-sequence companion, regardless whether the accretor is a white dwarf or 
a black hole. The fact that the orbital period of MAXI\,J1659$-$152 falls in the middle of the period gap, therefore, indicates 
that its companion is not a normal H-rich unevolved main-sequence star. It must be a nuclearly evolved object, such as 
produced by the above mentioned models of Pylyser \&\ Savonije (1988) and Pfahl et al.\ (2003). 
We are sure that in MAXI\,J1659$-$152 mass transfer from the companion is taking place, because also during quiescence, at least after the outburst, there is some 
X-ray emission (see Sect.~\ref{intro}). This indicates that there is always an accretion disk present in the system, which implies that mass transfer 
continues during quiescence. The companion must thus be filling its Roche lobe. The outburst of the system is, therefore, 
most likely due to some disk instability mechanism (e.g., Lasota 2001), and not due to some mass-loss event from a 
companion that is not filling its Roche lobe.  

We estimate the initial mass of the companion star as follows.
For the companion star to become evolved in a Hubble time, its mass must at least have been 1\,M$_{\odot}$. On the other hand, in 
order for the system to evolve to a short orbital period when its companion is driving mass transfer (partly) by nuclear evolution, 
its companion star cannot have been more massive than about 1.5\,M$_{\odot}$. The reason for this is that, in order to have the orbital 
period decrease whilst the companion is evolving, the orbital expansion due to the mass transfer driven by the evolution of the companion 
(transfer from the less massive to the more massive star makes the orbit expand) must be more than compensated by the orbital 
shrinking due to the orbital angular momentum loss from the system by magnetic braking. In fact, this implies that, the system 
must have started out with an orbital period below the so-called `bifurcation period', which is between 0.5~days and 0.8~days, 
depending on the system parameters (see Pylyser \&\ Savonije 1988, Pfahl et al.\ 2003). 
Above this limiting period in systems with an evolving companion 
and magnetic braking, the orbital period increases in time, whereas below this critical period it will decrease in time, since 
the angular momentum loss by magnetic braking wins from the orbital expansion due to the companion's internal nuclear evolution. 
In order to have orbital angular momentum loss by magnetic braking, the companion star must have a convective envelope 
(Verbunt \&\ Zwaan 1981, Rappaport et al.\ 1983, Knigge et al.\ 2011). 
This implies for main-sequence stars, that the initial mass cannot have been 
larger than 1.5\,M$_{\odot}$, since above this mass main-sequence stars have radiative envelopes. The conclusion is, therefore, 
that the companion star in MAXI\,J1659$-$152 must have started out with a mass between 1.0 and 1.5\,M$_{\odot}$. For examples of 
the evolution of systems with decreasing orbital periods, and with initial companion masses of 1.0 or 1.5\,M$_{\odot}$ and a 
4.0\,M$_{\odot}$ black hole accretor, see the evolutionary sequences 50 and 55 (table 1) of Pylyser \&\ Savonije (1988). 
The sequences Z55 and A55 in this paper started out with a companion mass of 1.5\,M$_{\odot}$ and evolved to an orbital period of 
2.4\,hr, where the companion masses had become 0.19 and 0.17\,M$_{\odot}$, respectively, with a central H content of 0.21 
and 0.09, respectively. The ages of the systems are then about 4.6 and 5.7 billion years, respectively, whilst the mass transfer 
continues at a very low rate (below 10$^{-11}$\,M$_{\odot}$\,yr$^{-1}$). It appears that these sequences may well represent 
the evolution of MAXI\,J1659$-$152.

\subsubsection{Size of the X-ray source and of the absorbing object}
\label{size}

Both Kennea et al.\ (2011) and Yamaoka et al.\ (2012) derive a mass of the black hole of $M_{\rm BH}\simeq 2.2$--3.1\,M$_{\odot}$ and
$M_{\rm BH}\simeq 3.6$--8.0\,M$_{\odot}$, respectively, 
based on the minimum innermost disk radius from X-ray spectral fits being the innermost stable circular orbit (assuming a distance of $>$6.1\,kpc 
and 5.3--8.6\,kpc, respectively, and an inclination angle of 60$^{\circ}$--75$^{\circ}$).
Shaposhnikov et al.\ (2012), on the other hand, estimated $M_{\rm BH}=20$$\pm$3\,M$_{\odot}$ by using an empirical relation between 
low-frequency QPO and X-ray spectral shape. The difference in mass estimates could be explained by taking into account the spin of the black hole
(Kennea et al.\ 2011, Yamaoka et al.\ 2012). 

Because of the rather large uncertainty in possible black-hole masses, we
first assume that the black hole has a mass of $M_{\rm BH}\gtrsim 3$\,M$_{\odot}$.
Using the companion mass estimate from Sect.~4.2.1.1, this leads to a mass ratio of $q=M_2/M_{\rm BH}\lesssim 0.065$ (see also Kennea et al.\ 2011).
Again assuming the companion is Roche-lobe filling, we can use the relation between the orbital separation, $a$,
and $q$ (Eggleton 1983), to get $a\gtrsim 1.33$\,R$_{\odot}$.
The duration of the ingress and egress of the absorption dips (a few seconds) 
and the duration of the dip activity (up to 40\,min) provide an estimate of the
extent of the object being absorbed and the absorber itself, respectively (see, e.g., \mbox{Kuulkers} et al.\ 1998, and
references therein). 
Assuming the black-hole mass follows the distribution of known BHXB masses, i.e., taking $M_{\rm BH} = 8$\,M$_{\odot}$ 
(\"Ozel et al.\ 2010, Kreidberg et al.\ 2012), 
and following \mbox{Kuulkers} et al.\ (1998), we find that the upper limit on the extent of the
object being absorbed is about 0.002\,R$_{\odot}$ or 0.005\,R$_{\odot}$ (i.e., about 1000\,km and 3500\,km, respectively), 
if the absorbing medium corotates with the binary frame
or corotates with matter in the accretion disk, respectively. Similarly, the size of the absorbing medium is estimated to be
1.8\,R$_{\odot}$ and 6.0\,R$_{\odot}$, respectively. For lower black-hole masses the estimated sizes are somewhat smaller.
The size of the object being absorbed is considerably larger than the innermost disk radius 
($\sim$30--100\,km, Mu\~noz-Darias et al.\ 2011, Yamaoka et al.\ 2012). The X-ray emission thus clearly comes from the inner part of the accretion disk, which is 
possibly surrounded by a disk wind or corona. The fact that the absorption dips do not drop to zero intensity at minimum is consistent
with this: part of the disk wind or corona stays always visible. The estimated sizes of the absorbing medium
are large, i.e., $\sim$1--3 times the orbital separation. This is very unlikely, so we suggest the absorbing medium
to be indeed spread over the outer part of the accretion disk, possibly along the accretion disk rim or the region above that, 
as discussed in Sect.~\ref{D_absorptiondips}.

\subsection{Optical counterpart of MAXI\,J1659$-$152}
\label{optical}

Using the fact that for LMXB transients the outburst amplitude ($\Delta {\rm V} = {\rm V}_{\rm min} - {\rm V}_{\rm max}$) is related to the 
orbital period, $P_{\rm orb}$ (Shahbaz \&\ \mbox{Kuulkers} 1998), we derive that the expected outburst amplitude for MAXI\,J1659$-$152, is $\Delta {\rm V}=11.4\pm 0.8$. 
The observed optical magnitude during outburst maximum was, ${\rm V}_{\rm max}\simeq 16.5$ (see Russell et al.\ 2010, Kennea et al.\ 2011).
This gives an expected V-magnitude in quiescence: V$_{\rm min}=27.9\pm 0.8$ (see also Kennea et al.\ 2011).
Applying the correction for inclination on the observation of V$_{\rm max}$ (see Sect.~\ref{distance}, Eq.~\ref{MV}) 
this would lead to expected V-magnitudes between 27.5 and 26.2, for inclination angles
between 65$^{\circ}$ and 80$^{\circ}$, respectively. If true, it may therefore not be easy to find the optical counterpart 
in quiescence.

A Pan-STARRS 1 (PS1) 3Pi sky-survey
observation on 2010 June 19, however, revealed an optical source consistent with the position of MAXI\,J1659$-$152
with an AB magnitude of about 22.8 in the r$_{\rm P1}$-band. The source was not detected in the other
filters (Kong et al.\ 2010, Kong 2012). 
Another observation, on 2012 March 23, with the Canada France Hawaii Telescope (CFHT), showed the source at a magnitude 
r'$\simeq$23.7 (Kong 2012).

Pan-STARRS uses filters similar to the SDSS g',r',i', and z'
filters (plus y$_{\rm P1}$ and w$_{\rm P1}$ filter), described in detail by 
Tonry et al.\ (2012). The Pan-STARRS system, like SDSS, is an AB system (e.g.,
Frei \&\ Gunn 1994), so r$_{\rm P1}$ = r'(AB). To get an estimate of the
brightness of the PS1 candidate in the Johnson V-band, we use
the r$_{\rm P1}$ band detection and the g$_{\rm P1}$ and i$_{\rm P1}$ upper limits. The limiting
magnitudes of the PS1 3Pi sky-survey in the g$_{\rm P1}$ and i$_{\rm P1}$ bands are
estimated to be 23.24 and 22.59 mag, respectively (for a 5-$\sigma$ point
source per visit, e.g., Chambers 2006). 
The Pan-STARRS measurements thus lead to g$_{\rm P1}$$-$r$_{\rm P1}\gtrsim 0.44$ and r$_{\rm P1}$$-$i$_{\rm P1}\lesssim 0.21$. 
Using the conversions from the Pan-STARRS system to the Johnson system
(Tonry et al.\ 2012), we derive ${\rm V}\gtrsim 23.0$.
This is consistent with the lower limit based on the non-detection in the USNO-B catalogue, ${\rm V}>21$, see Kennea et al.\ (2011),
although it is off by about 3 magnitudes from that expected (${\rm V}\gtrsim 26.2$, see above).
Another possibility, although we regard it as unlikely, is that the Shahbaz \&\ \mbox{Kuulkers} (1998) relation 
breaks down at short orbital periods. 

An M5V star in the Pan-STARRS system gives g$_{\rm P1}$$-$r$_{\rm P1}\simeq 1.2$ and r$_{\rm P1}$$-$i$_{\rm P1}\simeq 1.4$
(J.~Tonry 2012, priv.\ comm.).
Assuming that in quiescence the optical contribution solely comes from the companion star, the observed value of
r$_{\rm P1}$$-$i$_{\rm P1}$ of the optical candidate is not compatible with a M5V star. Instead, the relationship between
r$_{\rm P1}$$-$i$_{\rm P1}$ and spectral type suggests the candidate to be of a type earlier than about G3
(J.~Tonry 2012, priv.\ comm.).
Our assumption above may, however, not be correct. The accretion disk can in quiescence still contribute significantly
(e.g., Jonker et al.\ 2012, and references therein). Assuming a disk contribution of about 50\%, the discrepancy between
the expected brightness of the companion and the suggested optical star becomes about one magnitude less, but is not enough to
solve the total difference of about three magnitudes. Also, the quiescent disk in late-type, short-period BHXBs are not expected to be so hot ($\sim$6000\,K) 
that the total optical emission mimics a G3-type star. The latter can also not be alleviated by the reddening towards
MAXI\,J1659$-$152, which is only moderate (see Sect.~\ref{distance}).

The source reported by Kong et al.\ (2010) and Kong (2012) seems brighter than expected, which cannot be
explained by the inclination (see above), a disk contribution or reddening (see Sect.~\ref{distance}) alone, 
and the fact that not all colours are consistent with a M5V star, which can also not be solved by the presence of
an accretion disk. We, therefore, conclude that we still (Kuulkers et al.\ 2012a; see also Kennea et al.\ 2011) 
can not exclude the possibility that the optical source is a foreground star.

\subsection{Distance to MAXI\,J1659$-$152}
\label{distance}

The distance to MAXI\,J1659$-$152 can be estimated in various ways. 
At maxima during outbursts, the optical brightness is dominated by emission from the accretion disk.
Assuming that all of the optical flux in quiescence comes from the companion, we can use equation 5 of Shahbaz \&\ \mbox{Kuulkers} (1998)
to derive an estimate of the absolute disk brightness, which is then only a function of $P_{\rm orb}$.
This leads to a rough estimate of M$_{\rm V,disk}=1.0\pm 0.8$\,mag.
To estimate the interstellar reddening, $A_{\rm V}$, we use the relation from G\"uver \&\ \"Ozel (2009) between the hydrogen column density, $N_{\rm H}$, and $A_{\rm V}$. 
The measured values of $N_{\rm H}$ by the {\it Swift}/XRT during and after the outburst vary between 2.4 and $6\times 10^{21}$\,cm$^{-2}$ 
(Kennea et al.\ 2011, Yamaoka et al.\ 2012). This is slightly higher than the estimated Galactic H~I column density in the direction of MAXI\,J1659$-$152 using 
Kalberla et al.\ (2005), i.e., the weighted average $N_{\rm H}=1.74\times 10^{21}$\,cm$^{-2}$, indicating there is intrinsic absorption 
in the system (Kalamkar et al.\ 2011, Kennea et al.\ 2011). This is not unexpected, given the high inclination of the source (see Sect.~\ref{D_absorptiondips}).
To derive $A_{\rm V}$ we, therefore, use the estimated Galactic $N_{\rm H}$, which leads to $A_{\rm V}\simeq 0.8$.\footnote{Note that
this is different from the values quoted in D'Avanzo et al.\ (2010; $A_{\rm V}\simeq 0.34$), and
Kuulkers et al.\ (2012a) and Kaur et al.\ (2012; $A_{\rm V}\simeq 1.1$), based on the measured $N_{\rm H}$ from {\it Swift}/XRT 
observations reported by Kennea et al.\ (2010). Kennea et al.\ (2011) used $A_{\rm V}=1.85$ (in the UVOT photometric system), which is an 
upper limit based on a value of $E({\rm B}-{\rm V})$=0.606 in the direction of MAXI\,J1659$-$152.}
With the observed maximum V-band magnitude of V$_{\rm max}=16.5$ (see Sect.~\ref{optical}), 
using the distance modulus (see equation 10 of Shahbaz \&\ \mbox{Kuulkers} 1998), we then infer that the distance to
MAXI\,J1659$-$152 (see also Kennea et al.\ 2011) is 8.6\,kpc, with an estimated uncertainty of 3.7\,kpc.
Using this distance, we find a height above the Galactic plane, $z$, of 2.4$\pm$1.0\,kpc (see also Kennea et al.\ 2011).

As noted by Miller-Jones et al.\ (2011), our derived distance may be an overestimate, since the 
relation from Shahbaz \&\ \mbox{Kuulkers} (1998) does not take into account the inclination angle of the system.
At large inclination angles the projected area of the disk may be smaller, and, therefore, the 
apparent disk brightness be less. One thus has to correct the observed optical magnitude at maximum
for this effect. For an optically thick accretion disk this correction amounts to (Paczy\'nski \&\ Schwarzenberg-Czerny 1980, 
assuming a limb-darkening coefficient of 0.6, see also Warner 1987):
\begin{equation}
\label{MV}
\Delta M_{\rm V} = -2.5 \log{[(1+1.5\cos{i})\cos{i}]}
\end{equation}
For inclination angles between 65$^{\circ}$ and 80$^{\circ}$ we find that $\Delta M_{\rm V}$ varies from 0.40 to 1.65.
This leads to revised distance estimates between 7.1$\pm$3.0\,kpc and 4.0$\pm$1.7\,kpc, respectively.
Accordingly, $z$ then has values between 2.0$\pm$0.9\,kpc and 1.1$\pm$0.5\,kpc, respectively.

For a M5V star (Sect.~4.2.1.1) the absolute V-band magnitude is about 11.8 (e.g., Zombeck 1990).
Although not very constraining, the observed value of V for the proposed optical counterpart in quiescence
(see Sect.~\ref{optical}) translates to a lower limit to the distance of about 1.1\,kpc.
Assuming our expected V-band magnitude in quiescence (V$_{\rm min}=26.2$--27.9~mag; 
Sect.~\ref{optical}) we derive a distance of 5.3--8.7\,kpc from the distance modulus.

Using the PS1 optical counterpart in quiescence proposed by Kong et al.\ (2010, Sect.\ 4.3), and assuming that 
the companion is the sole contributor during optical quiescence, Miller-Jones et al.\ (2011) estimated the distance to
MAXI\,J1659$-$152 to be 1.6--4.2\,kpc. 
Assuming the companion to be an M5V (see Sect.~4.2.1) or an M2V (Jonker et al.\ 2012) type star, and using the
CFHT optical detection (see Sect.~4.3), Kong (2012) derived distances of 2.3--3.8\,kpc and 4.6--7.5\.kpc, respectively.
Kaur et al.\ (2012) estimate a lower limit of 4$\pm$1\,kpc, based on the measured 
radial velocity distribution of the interstellar Na~{\sc I}~D and Ca~{\sc II}~H\&K lines.
Using the relation between the absolute magnitude of a LMXB as a function of the orbital period and the X-ray luminosity
(van Paradijs \&\ McClintock 1994), and assuming that MAXI\,J1659$-$152 reached 10\%\ of the Eddington luminosity
at maximum of the outburst, Kennea et al.\ (2011) derived distances in the range $\simeq$3.2--7.5\,kpc for values of 
$A_{\rm V}$ from 1.85 to 0. Assuming $M_{\rm BH}>3.2$\,M$_{\odot}$ and that MAXI\,J1659$-$152 at the peak of the outburst radiates at more
than 10\%\ of the Eddington limit, they derive a distance $>$6.1\,kpc.
Jonker et al.\ (2012), using a M2V type companion star and an accretion disk contribution to the 
quiescent optical light of 50\%, derive a distance of 5.9\,kpc with an estimated uncertainty of 2\,kpc.
Shaposhnikov et al.\ (2012), using their spectral-timing correlation scaling method, found an upper limit of 7.6$\pm$1.1\,kpc, whereas  
Yamaoka et al.\ (2012) estimate an upper limit of about 8.6\,kpc. The latter authors combined various information, including the 
fact that the soft-to-hard transition in BHXB transients occurs at 1--4\%\ of the Eddington luminosity (Maccarone 2003; see also Sect.~\ref{transitionluminosity}).
None of these estimates are particularly robust, however, and there is a considerable spread.
Our initial estimates in the beginning of this subsection are consistent with the above quoted values; therefore, in the rest of this paper we adhere to our values of the 
distance of 8.6\,kpc, and the distance above the Galactic plane of $z=2.2$\,kpc.

\subsection{On the soft to hard state transition luminosity and the maximum outburst luminosity}
\label{transitionluminosity}

Black-hole transient sources transit from the soft state to the hard state when the source luminosity is about 1--4\%\ of the 
Eddington luminosity, $L_{\rm Edd}$, with a mean value of 1.9$\pm$0.2\%\ (Maccarone 2003; see also Dunn et al.\ 2010 for a discussion). 
Both the spectral and timing behaviour can be used to determine the exact time of transition (e.g., Dunn et al.\ 2010, Belloni 2010, 
Mu\~noz-Darias et al.\ 2011a, and references therein). 
In the hard state, power-law emission dominates and the power-density spectrum shows a strong noise component.
MAXI\,J1659$-$152 reached the hard state between day 39.1 and day 40.1 (Mu\~noz-Darias et al.\ 2011b).
Maccarone (2003) used a cut-off power-law spectrum, with spectral index, $\Gamma=1.8$, and cut-off energy of 200\,keV, integrated between
0.5\,keV and 10\,MeV to derive the bolometric correction. To estimate the bolometric flux near the transition of the soft-to-hard state
for MAXI\,J1659$-$152, we used the spectrum observed on day 41.0 from Mu\~noz-Darias et al.\ (2011b), which showed 
$\Gamma=1.80^{+0.02}_{-0.03}$.\footnote{Note that Yamaoka et al.\ (2012) used
the estimated bolometric flux from an observation of the hard state after the state transition, on day 44.1.
Their observation is about 3~days later
than the observation we take as being near the transition; the flux had declined by about 20\%\ in that time.}
We used their spectral parameters, added a cut-off at 200\,keV 
(which is well outside the RXTE/PCA energy range), and then integrated between 0.5\,keV and 10\,MeV. 
To estimate the uncertainty in the bolometric flux we randomised the spectral parameters
using the fit values (and a fixed high-energy cut-off at 200\,keV) and their derived 1$\sigma$-errors from Mu\~noz-Darias et al.\ (2011b). 
This was done 100\,000 times and we recorded the resulting integrated fluxes between 0.5\,keV and 10\,MeV.
The flux distributions are significantly skewed towards larger fluxes. We, therefore, fitted the flux distributions
below and above the peak of the distribution with Gaussians with different widths. The plus and minus errors in
the integrated flux derived from the spectral fits were then taken as the 1$\sigma$ widths (see, e.g., Kuulkers et al.\ 2010a).
We find an estimated (unabsorbed) bolometric flux of 4.20$^{+0.42}_{-0.34}\times 10^{-9}$\,erg\,cm$^{-2}$\,s$^{-1}$.
Since the mass of the black hole can be assumed to be at least 3\,M$_{\odot}$, we derive an upper limit on the transition luminosity for
MAXI\,J1659$-$152 of about 10\%\,$L_{\rm Edd}$.
For a canonical black-hole mass of 8\,M$_{\odot}$ (\"Ozel et al.\ 2010, Kreidberg et al.\ 2012) we derive a transition luminosity of $\simeq$3.7\%\,$L_{\rm Edd}$.
The latter is consistent with the soft to hard state transition luminosity derived by Maccarone (2003), and would suggest a more canonical
mass of the black hole in MAXI\,J1659$-$152.
We note, however, that the black-hole transient GRO\,J1655$-$40 did not fit Maccarone's (2003) relation.
GRO\,J1655$-$40 is also a high-inclination, dipping source, similar to MAXI\,J1659$-$152 (see Kuulkers et al.\ 1998, 2000; Sect.~\ref{D_absorptiondips}). 
The above, and given the range in distance (Sect.~\ref{distance}) and black-hole mass (Sect.~\ref{size}) estimates, 
leads us to conclude that our test of Maccarone's (2003) relation should 
be used with some caution in the case of MAXI\,J1659$-$152.

The maximum observed 2--10\,keV flux during MAXI\,J1659$-$152's outburst ($\simeq$9$\times$10$^{-9}$\,erg\,cm$^{-2}$\,s$^{-1}$ on day 13.1; Kennea et al.\ 2011)
translates to a maximum 2--10\,keV luminosity of roughly 8$\times$10$^{37}$\,erg\,s$^{-1}$.
This value is not unusual for transient BHXBs (see, e.g., Dunn et al.\ 2010).
From the X-ray spectral fit results by Yamaoka et al.\ (2012; their model A), one can derive that the extrapolated
maximum, unabsorbed, 3--200\,keV flux occurred on day 13.0, i.e., $\simeq$10$^{-8}$\,erg\,cm$^{-2}$\,s$^{-1}$.
For a black-hole with minimum mass of 3\,M$_{\odot}$ this gives an upper limit on the maximum outburst 3--200\,keV 
luminosity, $L_{\rm peak}$, of about 23\%\,$L_{\rm Edd}$; for a black hole with 8\,M$_{\odot}$ this would lead to 
$L_{\rm peak}\simeq 8.5$\%\,$L_{\rm Edd}$. The latter value is more or less as expected from the observed
relation between the maximum 3--200\,keV luminosity and $P_{\rm orb}$ (Wu et al.\ 2010).

\subsection{A class of short-period BHXB transients at high Galactic latitudes?}
\label{class}

It is interesting to note (see also Kennea et al.\ 2011, Yamaoka et al.\ 2012) that the two transient BHXBs with the shortest orbital periods, 
MAXI\,J1659$-$152 and Swift\,J1753.5$-$0127, are both found at high Galactic latitudes (b$^{II}=16.5$$^{\circ}$ and b$^{II}=12.2$$^{\circ}$, respectively), as well as a third short-period transient BHXB: 
XTE\,J1118+480 (see Zurita et al.\ 2008, b$^{II}=62.3$$^{\circ}$). Given their distances, the corresponding heights above the Galactic plane are:  
$z = 2.4$\,kpc (see Sect.~\ref{distance}), 1.1\,kpc (Zurita et al.\ 2008) and 1.6\,kpc (Jonker \&\ Nelemans 2004), respectively. 
This can be compared with the observed Galactic $z$-distribution of the BHXBs, which has an rms $z$-value of 0.625\,kpc (Jonker \&\ Nelemans 2004). 
Recently, another candidate for such a binary has been put forward:
Swift\,J1357.2$-$0933 (Casares et al.\ 2011, see also Yamaoka et al.\ 2012; b$^{II}=51$$^{\circ}$). This system has a M4V star 
as the companion (Rau et al.\ 2011, Casares et al.\ 2011) and is subluminous in the radio (Sivakoff et al.\ 2011), similar
to MAXI\,J1659$-$152 (see below).
However, as noted in Sect.~\ref{D_absorptiondips}, the optical outburst amplitude of $\sim$6 mag (Rau et al.\ 2011) 
is not consistent with the short period suggested by Casares et al.\ (2011).

Swift\,J1753.5$-$0127 shares other similarities than only the short orbital period with MAXI\,J1659$-$152. 
The shape of the overall outburst light curve as seen by {\it Swift}/BAT (Soleri et al.\ 2010), is very similar to that seen for 
MAXI\,J1659$-$152 (see also Kennea et al.\ 2011). Swift\,J1753.5$-$0127, however, has not been seen to turn off again so far, i.e., pre-outburst 
15--50\,keV flux levels have not been reached. Moreover, after it had reached the lowest levels about 
200\,days after the start of the outburst, it has been seen to vary on years time scales between 
about 20\,mCrab and 125\,mCrab (see, e.g., figure 1 of Soleri et al.\ 2010).  It has also been active in the radio wavelengths at a 
lower level compared to when the source was in outburst (Soleri et al.\ 2010). Low-level soft X-ray and radio activity has been reported for 
MAXI\,J1659$-$152 (Kennea et al.\ 2011, Yang et al.\ 2011a, 2011b, Yang \&\ Wijnands 2011a, 2011b, Miller-Jones et al.\ 2011, Jonker et al.\ 2012).
In the hard, 15--50\,keV band MAXI\,J1659$-$152 is not (yet) detected, from about 65\,days after the start of the outburst up to now. 
But it may be detected later again, if it follows the same trend as that seen for
Swift\,J1753.5$-$0127. 
One can compare the main outburst light curve of MAXI\,J1659$-$152 also with that of the BHXB transient XTE\,J1859+226 
(Casella et al.\ 2004; see also Sect.~\ref{transitiondips_discussion}). A high, soft, flux 
period is followed by a jump to a low, hard, one. In XTE\,J1859+226 the main outburst lasted also for about a month, with day to day variations similar to that seen
for MAXI\,J1659$-$152. The orbital period of XTE\,J1859+226, however, is not as extreme: $\simeq$6.6\,hr (Corral-Santana et al.\ 2011).

Yamaoka et al.\ (2012) argue that MAXI\,J1659$-$152 is a runaway micro-quasar, similar to XTE\,J1118+480, i.e., kicked out of the Galactic plane into the halo. 
We consider this indeed a plausible possibility for explaining the apparently large distances of these short-period systems 
from the Galactic plane for the following reasons (we say here "apparently", because the distances to these systems are still quite 
uncertain, causing considerable uncertainty in their distances to the Galactic plane). The available evidence for kicks imparted to 
the black holes in their formation events suggests that a sizeable fraction of black holes may receive rather large kick velocities 
at birth: GRO\,J1655$-$40 has an observed excess radial velocity relative to its local rest frame of 112$\pm$18\,km\,s$^{-1}$, and 
simulations of its evolution and formation by Willems et al.\ (2005), including the effects of kicks, indicate that the most likely 
kick velocity imparted in its formation was 105\,km\,s$^{-1}$. Similarly, XTE\,J1118+480 has an excess radial velocity of 
145$\pm$25\,km\,s$^{-1}$ and simulations of its evolution and formation by Fragos et al.\ (2007) indicate that a most likely kick 
velocity of about 200\,km\,s$^{-1}$ imparted in its formation event. On the other hand, simulations of the evolution of the 
Cyg\,X-1 system by Wong et al.\ (2012) indicate that the kick imparted at its formation was less than 77\,km\,s$^{-1}$ and most probably 
not more than about 40\,km\,s$^{-1}$ (see also Reid et al.\ 2011). 
The BHXB V404\,Cyg has a peculiar velocity of about 40\,km\,s$^{-1}$ (Miller-Jones et al.\ 2009),
indicating that its black hole did not receive a velocity kick larger than this value.
If many black holes would receive a velocity kick of the same order as those of GRO\,J1655$-$40 and XTE\,J1118+480, i.e., about 
100 to 200\,km\,s$^{-1}$, one may expect that many of the systems with short orbital periods will have high runaway velocities. 
As mentioned in Sect.~\ref{mandr}, in order for a BHXB to evolve to a very short orbital period, the initial mass of the companion 
star must have been $\lesssim$1.5\,M$_{\odot}$. If a typical black hole of mass 6 to 10\,M$_{\odot}$ received a kick of 
100--200\,km\,s$^{-1}$, the systems which have to drag along only a companion of low mass ($\lesssim$1.5\,M$_{\odot}$) 
will get the highest space velocities. In that case one expects the systems with the largest space velocities --- and therefore the 
largest mean distances to the galactic plane --- to be found among the BHXBs with the shortest orbital periods. (However, since there 
are also some black holes that do not receive large kicks, one still would also expect some of the systems with very short orbital 
periods to be located not far from the Galactic plane). 

The alternative possibility, that the systems with the shortest orbital 
periods were kicked out of globular clusters, seems very unlikely.
Black holes will be born very early in the life of a globular cluster (within the first ten million years). They will then by 
gravitational interactions with the other cluster stars rapidly sink to the cluster core, 
where the most massive objects of the cluster will concentrate. Calculations (Kulkarni et al.\ 1993)
show that in clusters of high central density the rapid dynamical evolution of the black-hole population in the cluster core leads to the ejection of nearly all the 
black holes on a short time scale. Because of the virtual absence of low-mass stars in these dense cores, it is very unlikely that they will have been able to capture 
a low-mass star before they were kicked out. However, for clusters of intermediate density, these authors found that some black holes survive in the cluster, and that 
some of these surviving black holes could form a LMXB. However, these BHXBs will stay in the cluster, since during a later phase in the 
evolution of the cluster, such a massive object cannot be kicked out any more by dynamical interactions with the cluster stars, because these are now of much lower 
mass than the black hole. It thus seems virtually impossible that systems like MAXI\,J1659$-$152 were formed in a globular cluster.

\subsection{Summary}

We have presented here a very detailed analysis of the X-ray and UV light curves of MAXI\,J1659$-$152, obtained over $\sim$260 days in 2010--2011 
with {\it MAXI, RXTE, Swift} and {\it XMM-Newton}. Our analysis combined soft and hard energies and time resolutions to create a uniform presentation 
of the source intensity and timing behaviour. We identified two types of variations in the light curves, absorption and transition dips, 
characterised by differing spectral properties. The timing studies of these have led us to the following conclusions. 

\begin{itemize}

\item The absorption dips occur at the orbital period of the system and are due to the combined effects of high orbital inclination and obscuration 
by material from the companion star interacting with the accretion disk. The presence of these dips has allowed us to measure precisely the orbital 
period of the binary at 2.414$\pm$0.005\,hrs. This is the shortest BHXB period measured to date and is confirmed (Kuroda et al.\ 2010) 
with modulations of the optical light curve of the system. 

\item Using the absorption dips we have constrained the inclination of MAXI\,J1659$-$152 to be between $65^{\circ}$--80$^{\circ}$ and the spectral type of the companion star to be M5V. 

\item We also identified transition dips, which are most likely a result of instabilities in the inner accretion flow. During these dips the source 
softens with increasing intensity, in contrast to the absorption dips, which are less energy dependent with increasing count rates. The exact origin 
of these dips remains unknown.

\item Using an estimate of the black hole mass of $>$3\,M$_{\odot}$, and a ratio between the compact object and the companion mass of $q=M_2/M_{\rm BH}\lesssim 0.065$, we estimate the binary orbital separation to be $a\gtrsim 1.33$\,R$_{\odot}$.

\item The very short orbital period of the system allowed us to successfully argue that the companion is a nuclearly evolved star with initial mass of about
1.5\,M$_{\odot}$, which evolved to 0.19 (0.17)\,M$_{\odot}$ during 4.6 (5.7) billion years (depending on the evolutionary sequence). 

\item We adopt an inferred distance to the source of $8.6\pm3.7$\,kpc and a distance above the Galactic plane of $z=2.4\pm1.0$\,kpc, 
which translates to maximum 2--10\,keV and 3--200\,keV luminosities of $\sim$8$\times$$10^{37}$\,erg\,s$^{-1}$ and
$\sim$9$\times$$10^{37}$\,erg\,s$^{-1}$, respectively. 

\item There are by now three BHXB sources with relatively short periods and high Galactic scale heights. These have been argued to be runaway 
micro-quasars, i.e., systems kicked out of the plane into the halo. We argue that this hypothesis is better supported by the properties of 
MAXI\,J1659$-$152, when contrasted with the suggestion that the system was formed in a globular cluster. 

\end{itemize}

Future X-ray spectral analysis, as well as additional multi-wavelength studies, both in outburst and quiescence, will 
elucidate the shortest period BHXB further.

\begin{acknowledgements}
Partly based on observations obtained with {\it XMM-Newton}, an ESA science mission with instruments and contributions directly funded by ESA Member States and NASA.
This research has made use of data obtained through the High Energy Astrophysics Science Archive Research Center Online Service, provided by the NASA/Goddard Space Flight Center.
The MAXI/GSC data are provided by RIKEN, JAXA and the MAXI team, whilst the {\it Swift}/BAT transient monitor results are provided by the {\it Swift}/BAT team.
We especially thank the {\it XMM-Newton} Science Operations Centre for their prompt scheduling of the Target of Opportunity observations,
5\,hrs between trigger and start of observation on September 27!
We would also like to thank the {\it Swift} and RXTE teams for their scheduling of the many monitoring observations.
The research leading to these results has received funding from the European Community's Seventh Framework Programme 
(FP7/2007/2013) under grant agreement number ITN 215212 "Black Hole Universe". 
TMB acknowledges support to ASI-INAF grant I/009/10/0, as well as funding via an EU Marie Curie Intra-European Fellowship under contract no.\ 2011-301355.
EK thanks John Tonry for discussions regarding the Pan-STARRS~1 optical candidate, 
Vik Dhillon for supplying the `PERIOD' analysis package, which we used in our periodicity analysis,
and Kazutaka Yamaoka for providing the estimated 3--200\,keV fluxes from RXTE spectral fits.
EK and TMD acknowledge Sara Motta for her comments on the RXTE spectral analysis.
EK and MDT thank Roberto Vio for a discussion on the periodograms.
JC was supported by ESA-PRODEX contract N: 90057.
Last but not least, we thank the referee for his/her careful reading of the manuscript.
\end{acknowledgements}

\end{document}